\documentclass[lettersize,journal]{IEEEtran}
\usepackage{amsmath,amsfonts}
\usepackage{algorithmic}
\usepackage{algorithm}
\usepackage{array}
\usepackage{textcomp}
\usepackage{stfloats}
\usepackage{url}
\usepackage{verbatim}
\usepackage{graphicx}
\usepackage{cite}

\usepackage{subcaption}
\usepackage{makecell}
\usepackage{array}

\begin{document}

\title{Coherence in Control: Bridging Many-Core Mapping and Routing through Cost Unification}

\author{
Guochu Xiong,
Xiangzhong Luo,
and Weichen Liu%
\thanks{
Guochu Xiong and Weichen Liu are with the College of Computing and Data Science, Nanyang Technological University, Singapore. Xiangzhong Luo is with the School of Computer Science and Engineering, Southeast University, China. 
}
\thanks{
Corresponding author: Weichen Liu (email: liu@ntu.edu.sg).
}
\thanks{
This work is supported by the Ministry of Education, Singapore, under its Academic Research Fund Tier 2 (MOE-T2EP20224-0006).
}
}




\maketitle

\begin{abstract}
The rapid growth of data-intensive applications increases communication demands in many-core systems, where cache coherence, while essential for correct communication and data consistency, introduces substantial overhead due to frequent data sharing and coherence activities. As system scale and workload complexity grow, the resulting coherence traffic intensifies communication pressure, making the co-optimization of task mapping and routing essential for improving system performance. However, most existing approaches overlook cache coherence, leaving a substantial portion of coherence-induced communication unaccounted for and creating a mismatch between optimization objectives and actual communication patterns. Furthermore, by employing separate cost evaluators for mapping and routing, these approaches complicate objective coordination, may lead to conflicting decisions, and fail to capture the coherence-induced coupling between the two stages. To address these challenges, we propose CoCo, a coherence-aware co-optimization framework that jointly integrates task mapping and routing under a unified cost model for realistic scenarios. This unified model integrates communication cost, coherence overhead, and load imbalance into a single objective, enabling coherence-aware decision-making and effective trade-offs among optimization goals. Guided by this model, CoCo combines coherence-guided task mapping with reinforcement learning-based routing, where directional link weights are adjusted according to communication behavior to improve traffic distribution, enabling coherence-aware co-optimization for many-core systems. Experimental results show that CoCo reduces link utilization by 88.46\%, packet delay by 17.40\%, and execution time by 17.58\% compared with existing approaches, highlighting the importance of cache coherence in co-optimization design.


\end{abstract}


\section{Introduction}
\label{introduction}


With the rapid advancement of AI, data-intensive computing applications such as large-scale AI training~\cite{b22,b23} and parallel computing~\cite{b24} have become increasingly prevalent. These applications involve intensive data sharing and synchronization among processing elements, leading to substantial communication demands in many-core systems. To ensure correct execution in such environments, cache coherence mechanisms are required to maintain data consistency across distributed caches through operations such as invalidations and cache-to-cache transfers~\cite{b3,b16}, which generate significant coherence traffic over the on-chip network. As system scale and workload complexity continue to grow, this increasing coherence traffic places growing pressure on communication efficiency~\cite{b01}. Since communication behavior is jointly influenced by task placement and routing decisions, mapping–routing co-optimization~\cite{b1,b20,b21} has emerged as an important design approach to balance traffic distribution and shorten communication paths, thereby improving overall system performance.

\begin{figure}[htbp]
    \centering
        \includegraphics[width=1.02\linewidth]{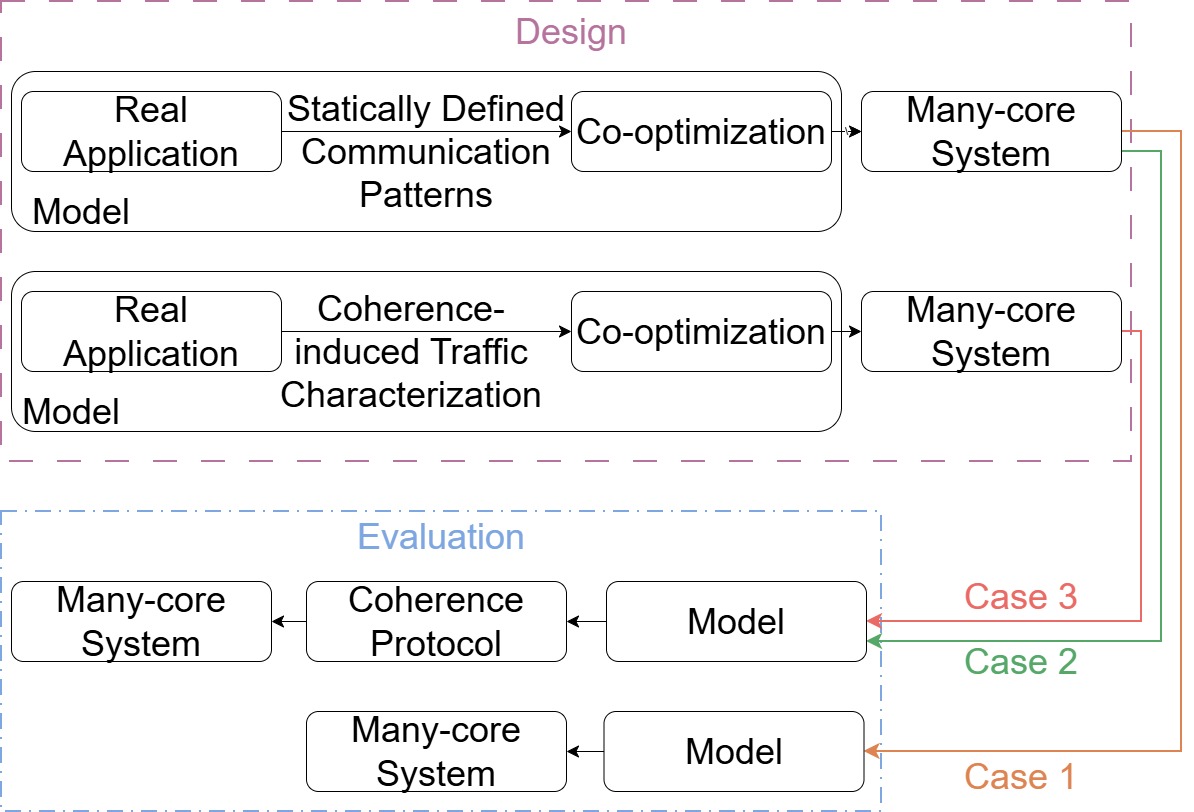}
    \caption{Difference cases for co-optimization designs}
    \label{fig:different_case}
\end{figure}

However, these methods typically suffer from two key limitations. First, existing co-optimization approaches overlook cache coherence effects. In realistic many-core systems, cache coherence is essential for maintaining data consistency across distributed caches and introduces additional communication through operations such as invalidations and data transfers. However, most prior designs~\cite{b1,b20,b21} perform co-optimization based on statically defined communication patterns that do not account for coherence-induced interactions between tasks, as illustrated in Fig.~\ref{fig:different_case}.
Specifically, in Case 1, both the design and evaluation stages assume a non-coherent communication model, resulting in simplified traffic patterns that fail to reflect realistic system behavior. In Case 2, although evaluation is conducted under a coherence-enabled system, the design phase still relies on non-coherent communication assumptions. As a result, coherence-induced traffic is not considered during optimization, leading to a mismatch between the optimization objectives and the actual communication behavior.

\begin{figure}[htbp]
    \centering
        \includegraphics[width=1.01\linewidth]{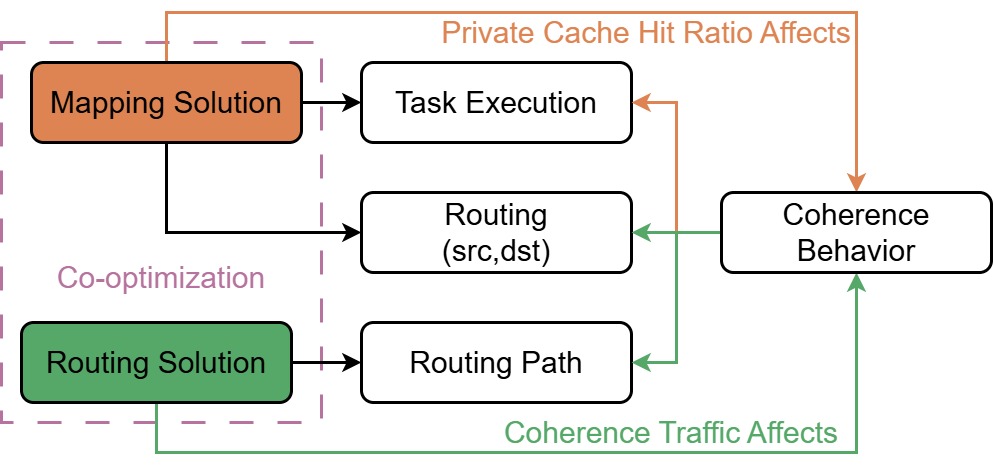}
    \caption{Interplay between task mapping, routing, and coherence behavior}
    \label{fig:relationship}
\end{figure}

To address these limitations, Case 3 incorporates cache coherence into both the design and evaluation stages, enabling accurate communication modeling. As illustrated in Fig.~\ref{fig:relationship}, coherence behavior directly influences co-optimization by shaping both communication patterns and traffic distribution. From the mapping perspective, task placement determines the spatial distribution of data accesses and thus influences communication locality. When interacting tasks are mapped to nearby cores, shared data can be served from local caches, improving cache hit ratios and reducing coherence operations. Conversely, poor locality leads to more frequent coherence actions, such as invalidations, generating additional coherence traffic across cores. From the routing perspective, coherence operations introduce additional communication traffic that must be efficiently distributed across the network. Routing strategies therefore play a critical role in determining whether such traffic is balanced or leads to congestion. Inefficient routing may exacerbate congestion caused by coherence communication, while effective routing can mitigate its impact~\cite{b15}. Thus, by incorporating cache coherence through coherence-induced traffic characterization, Case 3 facilitates co-optimization strategies that explicitly capture the interaction among coherence, mapping, and routing, enabling optimization decisions that more accurately reflect realistic communication dynamics.

Second, existing co-optimization methods often employ separate cost evaluators for mapping and routing, as they target different aspects of communication behavior~\cite{b11,b9,b33}. Although this enables independent optimization of each component, it introduces two key challenges: (i) separating the evaluators can lead to inconsistent or conflicting decisions while increasing algorithmic complexity and complicating the trade-off between different optimization objectives, ultimately degrading overall system performance, and (ii) these methods fail to explicitly model cache coherence, despite the fact that mapping and routing are tightly coupled through coherence-induced communication, leaving a critical source of overhead unaccounted for.

To address these gaps, we propose CoCo, a coherence-aware co-optimization framework that jointly integrates task mapping and routing under a unified cost model in realistic scenarios. Our approach is motivated by three key observations: (i) Existing co-optimization designs often rely on statically defined communication patterns and dependencies, without explicitly incorporating cache coherence effects. However, in realistic many-core systems, cache coherence introduces additional communication through coherence activities. Thus, a mismatch is introduced between optimization objectives and actual runtime behavior, ultimately resulting in suboptimal performance. (ii) Task mapping and routing interact closely with cache-coherence behavior: task placement and routing choices determine the amount and distribution of coherence traffic, while the resulting coherence-induced contention and delays affect the effectiveness of these decisions. (iii) Existing approaches typically employ independent cost models for mapping and routing, which not only complicates the coordination of multiple optimization objectives and leads to potentially conflicting decisions, but also fails to account for the coherence-induced interdependence between the two, leaving a significant source of communication overhead insufficiently addressed.




Building on these insights, our contributions are as follows. (i) We introduce a unified coherence-aware cost evaluator model that integrates communication cost, coherence overhead, and load imbalance into a single objective. This formulation provides a consistent optimization criterion for both mapping and routing, balances the trade-off among different optimization objectives, and enables coherence-aware decisions throughout the co-optimization process. (ii) Guided by the unified cost model, we develop a coherence-guided task mapping method (CGTM) that incorporates coherence-induced communication behavior into the mapping process. By leveraging coherence traffic characteristics, CGTM places frequently interacting tasks closer together to enhance communication locality and reduce coherence-related contention, enabling more effective mapping under practical workloads. (iii) Based on the same cost model, we further develop a reinforcement learning-based directionally weighted routing scheme (RLDWR), which selects routing directions and adjusts directional link weights according to communication behavior to improve traffic distribution. Combined with CGTM, this design provides fine-grained routing control and enables effective co-optimization of mapping and routing under coherence-induced communication patterns. We evaluate the proposed framework using PARSEC benchmarks, demonstrating superior performance and efficiency compared with existing strategies.

\section{Related Work}
To the best of our knowledge, this is the first cache-coherence-aware co-optimization framework that jointly considers mapping, routing, and cache coherence within a single design targeting realistic applications.

Recent research has increasingly explored co-optimization as an effective approach to improving communication efficiency and overall system performance. SMART~\cite{b20} introduces a power-gating–aware mapping and routing scheme that reduces latency by minimizing wake-ups and hop count. Building on SMART, the SMT-based mapping approach~\cite{b21} applies an SMT framework to generate contention-free mappings and diversified routing paths. MARCO~\cite{b1} further introduces a communication-aware mapping and routing strategy that leverages the flexibility of interconnects to improve traffic distribution and reduce congestion.


Despite these advances, existing co-optimization methods still rely on separate cost models for mapping and routing. From the perspective of communication efficiency, mapping and routing techniques address different aspects of system behavior. Communication-aware mapping~\cite{b11} optimizes task placement based on communication intensity and congestion estimation, while HyDra~\cite{b308} combines design-time analysis with runtime adaptation to improve placement decisions. These approaches primarily capture the spatial characteristics of communication by determining where traffic is generated. In contrast, communication-aware routing approaches, such as Dyad~\cite{b33}, adapt routing paths based on runtime congestion indicators (e.g., queue lengths), while reinforcement learning-based methods~\cite{b29,b30} dynamically balance traffic by adapting to network states and structural characteristics. These methods focus on how traffic is distributed across the network.
Although both mapping and routing aim to improve communication efficiency, they are typically optimized using separate cost formulations that capture different aspects of system behavior. As a result, their decisions are not consistently aligned when combined, limiting the effectiveness of co-optimization and increasing optimization complexity. This limitation highlights the need for a unified cost formulation that can consistently guide both mapping and routing.

In parallel, these co-optimization methods rely on statically defined communication patterns and dependencies, largely overlook cache coherence as a key determinant of system efficiency. Several works have attempted to incorporate coherence, but only partially. On the mapping side, Martin Rapp et al.~\cite{b302} consider last-level cache latency, and H. Ding et al.~\cite{b303} model shared-cache conflicts, yet neither captures coherence-induced data sharing in realistic workloads. On the routing side, a DRL-based coherence-aware strategy~\cite{b15} incorporates coherence into routing decisions. However, it does not consider task placement, leaving mapping unoptimized and resulting in suboptimal system-level performance.



\section{Methodology}
\label{method}

\begin{figure}[htbp]
    \centering
        \includegraphics[width=\linewidth]{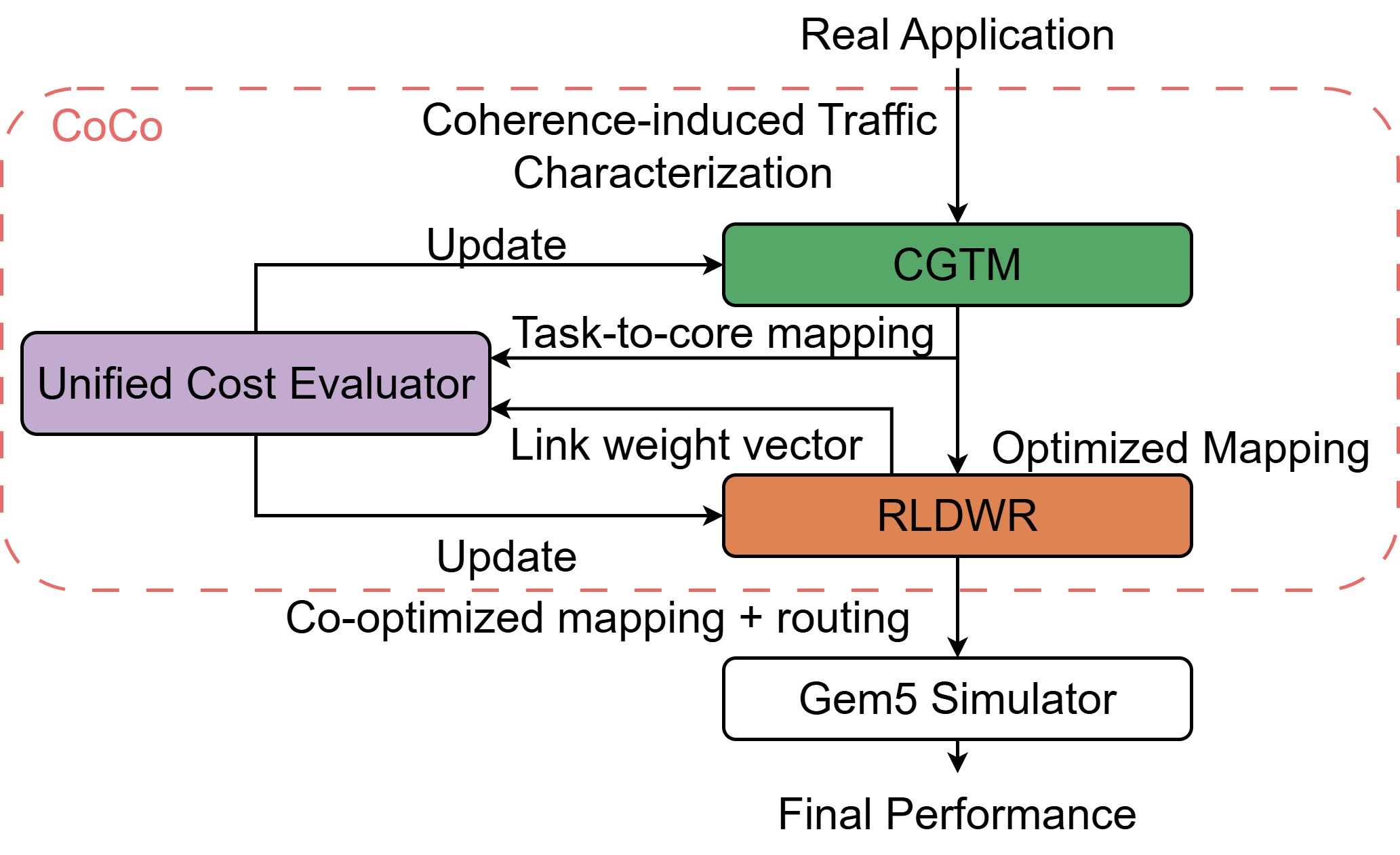}
    \caption{Workflow of CoCo}
    \label{fig:workflow}
\end{figure}

In this section, we present CoCo, a coherence-aware co-optimization framework illustrated in Fig.~\ref{fig:workflow}. CoCo operates in three key phases: (i) Coherence-Guided Task Mapping (CGTM), which incorporates coherence-induced traffic characterization into mapping decisions through a Tabu-guided exploration process, enabling the mapping procedure to capture communication impacts arising from the interaction between task placement and coherence behavior beyond conventional distance-based metrics. (ii) Reinforcement Learning (RL)-based Directionally Weighted Routing (RLDWR), where directional link weights are iteratively adjusted through learned actions during training to encode routing preferences according to communication behavior, enabling communication-efficient routing under coherence-induced communication patterns. (iii) a unified cost evaluator model that integrates coherence behavior, communication cost, and load balance into a single objective. Guided by this evaluator, CoCo jointly optimizes task mapping and routing, allowing the two processes to co-evolve toward communication-efficient configurations rather than being optimized independently. 

The overall optimization process proceeds as follows. CoCo evaluates each mapping–routing pair using the unified cost evaluator (Section~\ref{cost}). In the mapping stage (Section~\ref{mapping}), candidate task placements are iteratively explored through pairwise task swaps following a Tabu-guided search, and each candidate is evaluated using the unified cost. During this process, coherence-induced communication is incorporated into the cost evaluation, allowing the mapping to account for realistic communication behavior. For each candidate mapping, directional link weights are randomly sampled to approximate routing behavior, and the resulting mapping–routing pair is evaluated using the unified cost. The mapping with the lowest cost encountered during the search is selected as the final task placement $\mathcal{M}$. After determining the final mapping $\mathcal{M}$, CoCo fixes the task-to-core assignment and performs routing optimization (Section~\ref{routing}). Routing is refined through an offline reinforcement learning (RL) process that iteratively updates the directional weight vector $\mathbf{w}$. At each step, an action selects a direction and adjusts its corresponding weight, producing a new vector $\mathbf{w'}$, which is evaluated using the unified cost under $(\mathcal{M}, \mathbf{w'})$. The reward guides the update process toward weight configurations that reduce coherence-induced communication overhead. Across training episodes, the directional weights progressively improve and converge to an optimized vector $\mathbf{w}^*$.
Together, the selected mapping $\mathcal{M}$ and learned routing weights $\mathbf{w}^*$ constitute CoCo’s final coherence-aware mapping–routing solution, enabling effective optimization in realistic scenarios. This tight coupling between placement and routing represents the core novelty of CoCo. The following subsections describe each phase in detail.


\subsection{Coherence-Guided Task Mapping}
\label{mapping}
Existing studies~\cite{b10,b8,b25,b11} have explored a range of task-mapping algorithms, including tabu search, simulated annealing, and genetic algorithms, as well as approaches aimed at alleviating communication bottlenecks. Among these, tabu search provides an effective balance between solution quality and computational efficiency in task mapping. In parallel, communication-aware mapping techniques~\cite{b11,b308} demonstrate strong capability in improving communication efficiency by reducing network congestion. However, existing frameworks still rely on statically defined communication patterns and dependencies, without incorporating coherence interactions. As a result, they fail to capture realistic runtime communication behavior, particularly cache coherence traffic. Consequently, a substantial portion of coherence-induced communication is overlooked during the design phase, leading to a mismatch between optimization objectives and actual execution behavior. This discrepancy results in suboptimal designs and limits the applicability of existing approaches to realistic workloads.

To address this limitation, CoCo introduces Coherence-Guided Task Mapping (CGTM) with a routing-aware and coherence-aware assessment mechanism. By incorporating coherence-induced communication into the evaluation and jointly considering task placement and routing behavior through the unified cost, CGTM enables mapping decisions to better reflect realistic communication patterns. Meanwhile, the framework leverages the efficiency of Tabu search to effectively explore the mapping space, allowing CoCo to improve communication efficiency while maintaining high search quality.

Given a realistic application task graph $G=(V,E)$ that captures communication among tasks, with coherence-induced traffic characterization derived from the application explicitly incorporated as input, a task-to-core mapping $\mathcal{M}: V \rightarrow C$ assigns each task to a core. Starting from a random initialization, CGTM employs a Tabu-search-based local exploration to iteratively improve task placement. At each iteration, a set of candidate mappings is generated by swapping two tasks $u, v \in V$, producing an updated mapping $M'$ such that $M'(u) = M(v)$ and $M'(v) = M(u)$. Recently explored swaps are temporarily recorded in a tabu list to prevent cycling, enabling the search to efficiently explore new regions of the mapping space while avoiding redundant evaluations.

To evaluate candidate mappings during the search process, we introduce a routing-aware and coherence-aware assessment mechanism, which constitutes a key novelty of our approach. Specifically, each candidate mapping $M'$ is evaluated together with a randomly sampled directional-weight vector $\mathbf{w} = [w_E, w_W, w_N, w_S]$, reflecting potential routing preferences. Under the mapping $M'$, communicating tasks are assigned to cores, determining the source and destination of communication pairs, while the directional weights $\mathbf{w}$ guide routing decisions and influence the selection of paths $\text{path}(u,v)$ between them, thereby determining the communication path cost. In addition, the coherence overhead between tasks is estimated using coherence-induced traffic characterization derived from the application, capturing the frequency and cost of coherence interactions. These components are jointly incorporated into the unified cost ($C_{\text{total}}$) (Section~\ref{cost}), where coherence-induced communication is explicitly modeled to reflect realistic communication behavior. 


Minimizing this cost naturally favors mappings that place frequently interacting tasks closer together, reducing both communication distance and coherence overhead. Furthermore, by evaluating each mapping under diverse sampled directional-weight configurations, the method captures the interaction between task placement and routing behavior across varying communication conditions. Unlike conventional approaches that typically evaluate mappings using fixed hop-count models or static routing schemes~\cite{b20,b21,b1}, this design reduces sensitivity to any single routing assumption by considering multiple routing configurations during mapping evaluation. Consequently, the subsequent routing optimization stage refines routing decisions based on a routing-aware mapping solution, reducing the likelihood of a mismatch between the selected mapping and the final routing configuration.

Guided by $C_{\text{total}}$, the search iteratively selects improved mappings. To enhance efficiency, recently explored swaps are recorded to prevent immediate revisiting unless a better solution is identified. Among admissible candidates, the first mapping that reduces $C_{\text{total}}$ is accepted, enabling steady refinement toward communication-efficient solutions.


\subsection{RL-Based Directionally Weighted Routing}
\label{routing}

To improve communication efficiency and reduce network congestion, prior work~\cite{b9,b33} has explored routing strategies that adjust routing decisions according to observed communication patterns, commonly referred to as communication-aware routing. However, these approaches often incur high computational overhead, making them unsuitable for co-optimization scenarios. With the rapid development of AI techniques, reinforcement learning (RL)-based routing~\cite{b29,b30} has emerged as a promising alternative. Nevertheless, these methods typically lack awareness of coherence-induced communication dynamics. Although several studies~\cite{b15} have investigated incorporating coherence information into RL-based routing, these approaches are designed as standalone routing optimizations and cannot be directly integrated with task mapping. Consequently, enabling coherence-aware routing within a unified mapping–routing co-optimization framework remains a significant challenge. 

\begin{figure}[htbp]
    \centering
        \includegraphics[width=1.03\linewidth]{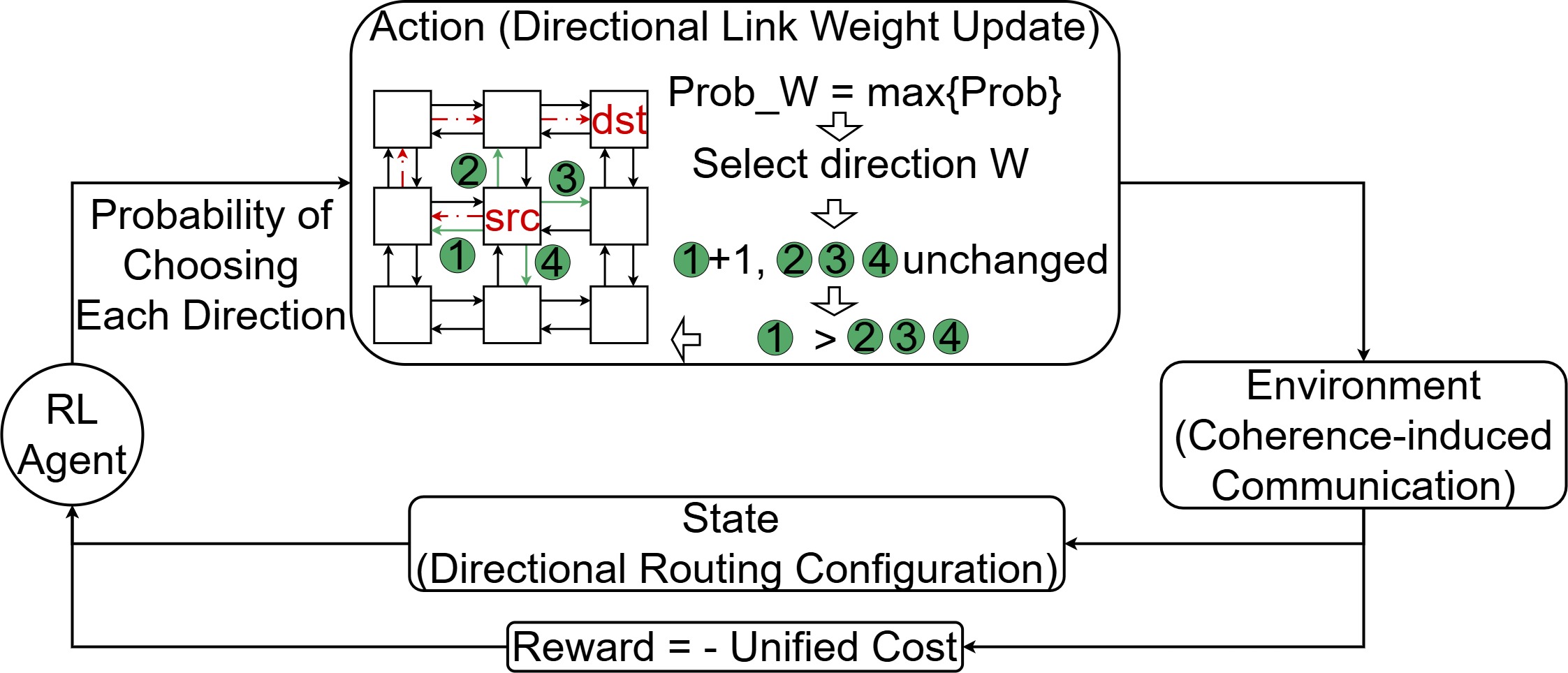}
    \caption{Workflow of RLDWR}
    \label{fig:workflow_routing}
\end{figure}

To address this challenge, CoCo proposes RL-Based Directionally Weighted Routing (RLDWR), an offline reinforcement-learning-based routing optimization framework that iteratively updates directional link weights according to communication and coherence characteristics, as illustrated in Fig.~\ref{fig:workflow_routing}.


\textbf{State.}
Instead of representing the state using global traffic metrics~\cite{b29,b30,b15}, our method encodes the current directional routing configuration as the state. Specifically, in a mesh topology, the state is defined as $S = [w_E, w_W, w_N, w_S]^T$, where each element corresponds to the routing weight assigned to a direction  (i.e., East, West, North, and South). These weights directly determine routing preferences when multiple output ports are available during packet forwarding. This constitutes a key novelty of our design: rather than encoding high-dimensional network conditions, the state is defined as a compact set of control parameters that directly govern routing behavior. This representation significantly reduces the dimensionality of the search space while enabling efficient and interpretable exploration of routing configurations that capture communication and coherence characteristics arising from task communication and cache coherence interactions.



\textbf{Action.}
The action in RLDWR consists of two key steps and is supported by Garnet, which provides bidirectional links~\cite{b305}, allowing distinct routing weights to be assigned to each direction.
(i) Direction selection. Given the current state $S = [w_E, w_W, w_N, w_S]$, the Advantage Actor--Critic (A2C) policy outputs a probability distribution over the four routing directions:
\begin{equation}
    \pi(a \mid S) = [P_E, P_W, P_N, P_S]
\end{equation}
where $\pi(a \mid S)$ denotes the policy function, i.e., the probability of selecting action $a$ under state $S$, and $P_d$ represents the probability of selecting direction $d \in \{E, W, N, S\}$. Based on this distribution, one direction is selected to determine which routing weight will be adjusted.
(ii) Link-weight update. Once the direction has been selected, the routing weight associated with that direction is incremented, whereas the weights of the other directions remain unchanged. As a result, when multiple output ports are available, routing decisions preferentially select directions with larger weights, enabling the optimized routing configuration to promote more communication-efficient paths.

The key novelty of RLDWR lies in using a compact directional-weight representation to optimize routing preferences. Compared with finer-grained routing policies, such as region-based~\cite{b02} or per-router routing policies~\cite{b03}, which emphasize localized routing optimization, the proposed design avoids a large number of routing-control variables and reduces the complexity of coordinating local routing decisions, thereby lowering training overhead and improving optimization efficiency. Consequently, the proposed global directional policy represents a deliberate tradeoff between routing flexibility and optimization complexity, enabling scalable optimization while maintaining compatibility with the mapping stage. Furthermore, since RLDWR only adjusts directional link weights used for route selection without modifying the underlying routing protocol or channel-dependency structure, the deadlock-avoidance properties of the underlying routing framework are preserved. This lightweight design also enables RLDWR to be readily incorporated into existing routing frameworks with minimal integration overhead, improving its practical deployability. As a result, RLDWR can efficiently identify communication-efficient routing configurations with lower coherence overhead.


\textbf{Reward.}
The reward in RLDWR is designed to explicitly capture the efficiency of coherence-induced communication:
\begin{equation}
    R = -C_{\text{total}}
\end{equation}
where $C_{\text{total}}$ is computed using the unified cost model described in Section~\ref{cost}. 
Unlike conventional routing approaches that rely on generic performance metrics~\cite{b9,b33,b29,b30}, RLDWR directly optimizes a coherence-aware communication objective, explicitly incorporating coherence effects into routing optimization. Moreover, although Multi-Objective Reinforcement Learning (MORL)-based~\cite{b2} can optimize multiple objectives simultaneously, it employs a multi-objective representation that is independent of the criterion used during mapping evaluation, potentially reducing coordination between the two stages and introducing optimization complexity. In contrast, CoCo adopts a unified cost model shared by both mapping and routing, allowing both stages to be guided by the same optimization objective. As a result, routing decisions remain consistent with the criterion used during mapping evaluation. Accordingly, the reinforcement learning process is formulated as an offline cost-minimization problem that searches for directional-weight configurations with lower communication cost and coherence overhead.




\textbf{Training.}
Routing optimization begins after the Coherence-Guided Task Mapping (CGTM) stage produces the final task-to-core assignment $\mathcal{M}$. During this phase, the mapping remains fixed to ensure that routing optimization is performed under stable communication patterns. The training process begins with an initial directional-weight vector, where each weight is randomly initialized. The directional routing configuration is then iteratively optimized through directional link-weight updates guided by the reward derived from the unified cost model. To ensure stable convergence, RLDWR continuously tracks the best directional-weight vector encountered during training. When a configuration with lower cost is observed, it is recorded and preserved as the current best solution. Exploration is controlled through a softmax temperature mechanism, which is gradually annealed to balance exploration and exploitation, allowing broader search in early stages and more stable updates as training progresses. Training terminates when either the preset episode limit is reached or the best directional-weight vector remains unchanged. The resulting directional-weight vector is then deployed as the routing configuration. During application execution, packets are routed using the selected directional weights, and no additional policy updates or retraining are performed.


\subsection{Unified Cost Evaluator} 
\label{cost}
To improve communication efficiency, both mapping and routing designs have been extensively studied. Communication-aware mapping~\cite{b11, b308} focuses on spatial optimization by assigning tasks based on communication intensity and estimated congestion, thereby optimizing placement according to static communication characteristics. In contrast, communication-aware routing~\cite{b9,b33} adapts routing paths based on runtime congestion indicators such as queue lengths, optimizing traffic distribution according to dynamic network conditions. Thus, these approaches improve communication efficiency from different perspectives and are guided by distinct optimization criteria. As a result, separate cost evaluation models for mapping and routing are typically adopted, even when advanced communication-aware mapping and routing strategies are co-optimized~\cite{b20,b21,b1}. This leads to two key challenges: (i) it complicates the coordination of optimization objectives, potentially resulting in conflicting decisions and increased algorithmic complexity; and (ii) it fails to capture the coherence-induced coupling between mapping and routing, leaving a critical source of communication overhead unaccounted for.

To address these limitations, CoCo introduces a unified cost evaluator that jointly models coherence-induced communication and system-level optimization factors within a single objective, enabling coordinated mapping and routing co-optimization. The unified cost consists of two components: coherence-induced communication cost $C_{\text{comm}}$ and additional penalty terms.

\textbf{Coherence-induced Communication Cost.} Two dominant sources of communication overhead are considered: (i) the communication path cost $C_h$, determined by path length and directional preferences, and (ii) the coherence overhead $T_{\text{coh}}$, arising from cache coherence interactions between tasks. These two factors capture complementary aspects of communication latency in many-core systems, reflecting spatial (path-related) and temporal (coherence-induced) effects. The communication path cost is computed based on direction-dependent per-hop costs. Specifically, each hop $h$ along a routing path is assigned a cost $C_h$, determined by its direction (e.g., east, west, north, south), with different directions associated with different weights to favor or penalize certain routing behaviors. The overall communication path cost is then obtained by accumulating these per-hop costs along the routing path. In this way, the path cost reflects the total cost of data movement across the network and directly influences routing decisions. In contrast, the coherence overhead $T_{\text{coh}}$ captures the temporal component, accounting for delays introduced by coherence transactions between tasks. This aspect is particularly important, as communication latency depends not only on distance but also on the frequency and patterns of data exchanges. Moreover, these two components are directly influenced by mapping and routing decisions. $C_h$ depends on both mapping and routing, where mapping determines inter-task distance and routing affects path efficiency, while the coherence overhead $T_{\text{coh}}$ is governed by mapping through communication locality and interaction frequency. Thus, by jointly modeling these two aspects, the unified cost captures the key controllable sources of communication latency and effectively guides task mapping and routing optimization toward improved communication efficiency:
\begin{equation}
C_{\text{comm}} = \sum_{(u,v)\in E} \left( 
\alpha \cdot \sum_{h \in \text{path}(u,v)} \hat{C_h} 
+ \beta \cdot \hat{T_{\text{coh}}}(u,v)
\right)
\end{equation}
where $(u,v)\in E$ denotes a communication edge in the task graph, and $\text{path}(u,v)$ represents the routing path between the cores hosting tasks $u$ and $v$ under the given mapping (Section~\ref{mapping}) and routing decisions (Section~\ref{routing}). $T_{\text{coh}}(u,v)$ represents the coherence-induced communication time between tasks $u$ and $v$, obtained from the task graph.



Since the communication path cost and coherence overhead originate from different domains and may exhibit substantially different numerical magnitudes, both terms are scaled prior to aggregation to ensure balanced contributions. Therefore, scaling factors are introduced to align the numerical scales of different cost components and map them to comparable ranges before aggregation~\cite{b27.1}. These scaling factors are selected according to the numerical characteristics and units of the corresponding metrics and are not updated during optimization. The resulting scaled terms are denoted by $\hat{C}_h$ and $\hat{T}_{\text{coh}}(u,v)$, respectively. The coefficients $\alpha$ and $\beta$ control the relative importance of communication path cost and coherence overhead, enabling the optimization process to balance routing efficiency against coherence-related communication effects.

\textbf{Penalty.} While the coherence-induced communication cost $C_{\text{comm}}$ captures the primary overhead between communicating tasks, it mainly reflects the edge-level communication cost determined by routing distance and coherence-related delay under a given mapping and routing configuration. However, communication efficiency is also affected by global communication effects, such as resource contention and uneven traffic distribution, which are not fully captured by edge-level evaluation. As a result, optimizing coherence-induced communication cost alone may still lead to suboptimal performance when communication demand is concentrated on specific cores or routing directions.

To address this limitation, CoCo introduces additional penalty terms. First, a load imbalance penalty $C_{\text{imb}}$ is incorporated to capture the uneven distribution of communication and computation across cores. When tasks are mapped such that certain cores handle disproportionately high communication or processing demand, the resulting concentration of traffic can increase contention and delay. By penalizing such imbalance, the optimization promotes a more uniform distribution of communication load, improving overall communication efficiency.

Second, a directional imbalance penalty $C_{\text{dir}}$ is introduced to regulate how communication traffic is distributed across routing directions. Since routing behavior is guided by directional link weights, excessive preference for certain directions may lead to overuse of specific links, resulting in communication bottlenecks. To mitigate this effect, CoCo evaluates the distribution of directional hops and penalizes imbalanced usage across directions, encouraging a more balanced utilization of routing resources.



To account for the different numerical ranges of the coherence-induced communication cost and penalty terms, the same scaling methodology is applied prior to aggregation into the unified objective. This ensures balanced contributions from different cost components while maintaining a consistent objective scale throughout optimization. The overall unified cost is defined as:
\begin{equation}
C_{\text{total}} = \hat{C}_{\text{comm}} + \lambda_1 \hat{C}_{\text{imb}} + \lambda_2 \hat{C}_{\text{dir}},
\end{equation}
where $\hat{C}_{\text{comm}}$, $\hat{C}_{\text{imb}}$, and $\hat{C}_{\text{dir}}$ denote the normalized coherence-induced communication cost, load imbalance penalty, and directional imbalance penalty, respectively, and $\lambda_1$ and $\lambda_2$ control their relative importance.

The use of scaling factors throughout the proposed cost formulation offers three advantages. First, it prevents large-magnitude components from dominating the unified objective, enabling the weighting coefficients to more effectively regulate the relative contributions of different cost components. Second, because the scaling factors are determined offline, the same scaling methodology can be applied across different applications and system configurations. Third, the reward function (Section~\ref{routing}) does not rely on continuously updated normalization values. Consequently, the reward scale remains stable during reinforcement learning, avoiding the non-stationarity associated with changing normalization factors and thereby improving training stability. With all cost components scaled to comparable ranges, the coefficients $\alpha$, $\beta$, $\lambda_1$, and $\lambda_2$ directly determine the relative importance of communication cost, coherence overhead, and global communication effects within the unified objective. This enables coordinated optimization of communication efficiency, coherence overhead, and system-level communication balance~\cite{b012}.

\section{Experiment}
\label{experiment}




Experiments are conducted using the PARSEC 2.1 benchmark suite~\cite{b12} on the Gem5 
simulator~\cite{b27} with Garnet~\cite{b305}, targeting realistic many-core scenarios. Energy consumption is measured with McPAT \cite{b28}.
As summarized in TABLE~\ref{tab:system_config}, our framework supports multiple coherence protocols. For evaluation, we employ a directory-based MESI protocol~\cite{b16}, which is widely used in both research and industry (e.g., CHI~\cite{b120}), ensuring broad applicability and accurate coherence modeling. Most experiments are performed on a 64-core system using an $8\times 8$ 2D mesh, while the scalability study also includes 16-core and 144-core systems.

\begin{table}[htbp]
\centering
\caption{Supporting Configuration \& Platform Parameters}
\vspace{-3pt}
\label{tab:system_config}
\small
\renewcommand{\arraystretch}{1.2}
\begin{tabular}{|p{3.5cm}|p{4.6cm}|}
\hline
\multicolumn{2}{|c|}{\textbf{Supporting Configuration}} \\ 
\hline
\textbf{Parameters} & \textbf{Specification} \\ 
\hline
Protocols & Directory-based MESI/MSI/MOESI \\
\hline
\multicolumn{2}{|c|}{\textbf{Platform Parameters}} \\
\hline
\textbf{Parameters} & \textbf{Specification} \\ 
\hline
System Architecture & NoC-based many-core systems \\
\hline
Frequency & 2 GHz \\
\hline
Flit Size & 128 bits \\
\hline
Virtual Channels per Port & 4 \\
\hline
Flow Control & Credit-based \\
\hline
L1D Cache & Private, 64 KB \\
\hline
L2 Cache & Shared, 2 MB \\
\hline
Main memory & 512 MB \\
\hline
Cache Coherence Protocol & Two-Level Directory-based MESI \\
\hline
Topology Type & 2D Mesh ($4\times4$, $8\times8$, $12\times12$) \\
\hline
\end{tabular}
\end{table}

In our evaluation, CoTAM~\cite{b17} and CCTA~\cite{b15} are jointly employed to construct the application task graph and extract coherence-induced traffic characterization, which serve as inputs to the proposed co-optimization framework (Section~\ref{method}). Since the characterization is derived from workload-specific coherence interactions rather than synthetic traffic assumptions, it captures the dominant coherence communication patterns exhibited during execution and provides a representative approximation of runtime coherence behavior. Consequently, the extracted coherence information can be effectively incorporated into both task mapping and routing optimization. To assess the effectiveness of our approach under diverse communication behaviors, four representative benchmarks are selected: Vips (memory-intensive image processing), Canneal (irregular access with high coherence pressure), Fluidanimate (communication-intensive scientific workload), and Blackscholes (compute-centric with low coherence demand). Together, these benchmarks cover a broad range of coherence-induced traffic patterns, enabling a comprehensive evaluation across realistic workload scenarios.

The evaluation focuses on three aspects: (i) The Coherence-Guided Task Mapping (CGTM) component is evaluated against representative mapping strategies under deterministic XY routing to assess its effectiveness and analyze the impact of cache coherence on task mapping. (ii) The Reinforcement Learning-based Directional Weight Routing (RLDWR) component is evaluated against representative routing approaches under a fixed mapping to assess its effectiveness in improving routing performance. (iii) The integration of mapping and routing is first analyzed by comparing three configurations: CGTM-only, RLDWR-only, and the full CoCo framework. Specifically, CGTM-only applies CGTM for task mapping with deterministic XY routing, while RLDWR-only employs a fixed mapping with RLDWR. This comparison demonstrates the benefits of co-optimization. Subsequently, CoCo is evaluated against state-of-the-art approaches, to assess its effectiveness in improving overall system performance, including SMART~\cite{b20}, which employs power-gating–aware task mapping with deterministic routing, the SMT-based approach~\cite{b21}, which constructs contention-free mappings with diversified routing paths, MARCO~\cite{b1}, which introduces a communication-aware mapping and routing strategy by leveraging flexible interconnects.



\subsection{Analysis of Coherence Traffic in NoC Communication}
\begin{figure}[htbp]
    \centering
        \includegraphics[width=0.95\linewidth]{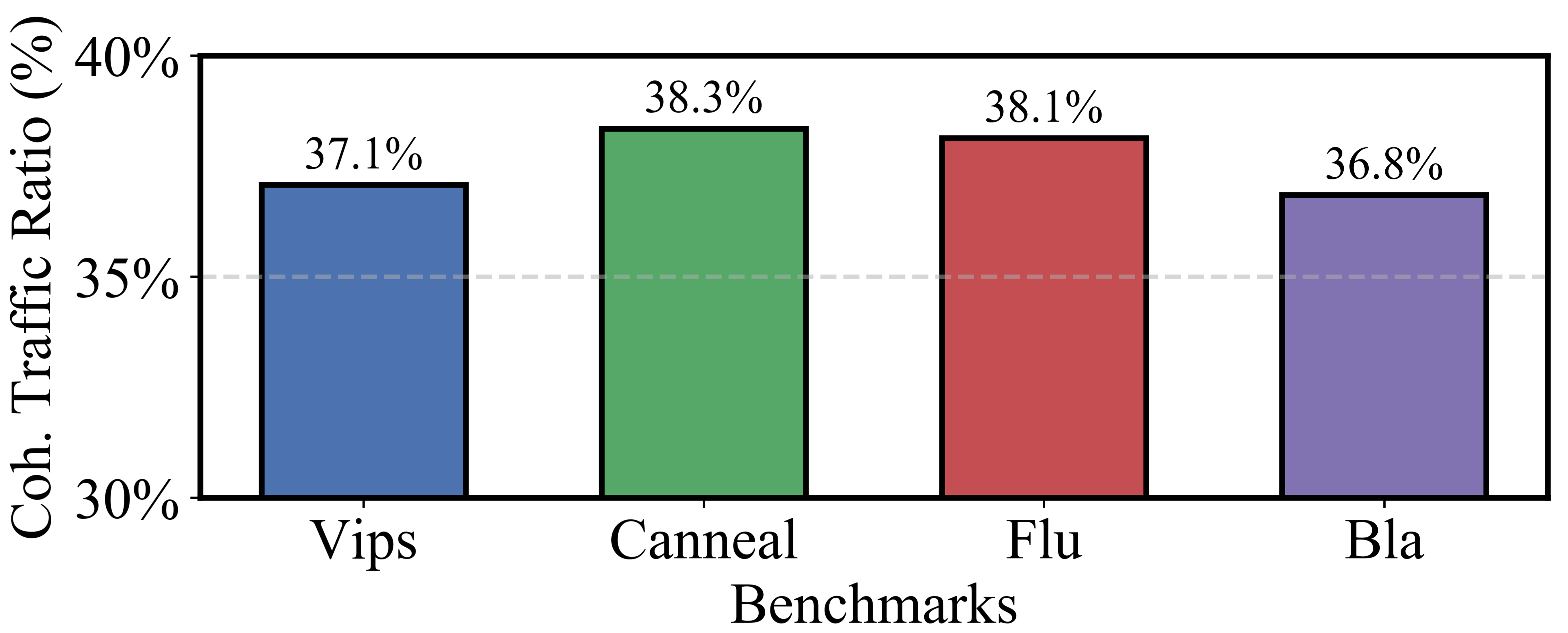}
   \caption{The ratio of coherence traffic to total NoC traffic}
    \label{fig:traffic-ratio}
\end{figure}

Before evaluating the performance of CoCo, we first quantify the contribution of cache-coherence traffic to overall NoC communication to support the motivation presented in Section~\ref{introduction}. As shown in Fig.~\ref{fig:traffic-ratio}, cache-coherence traffic accounts for a substantial portion of the total NoC traffic across all evaluated benchmarks, representing 37.1\%, 38.3\%, 38.1\%, and 36.8\% of the total network traffic for Vips, Canneal, Fluidanimate (Flu), and Blacksholes (Bla), respectively. These results indicate that coherence-induced communication constitutes a significant component of NoC traffic and can impose considerable communication overhead on the network. This finding motivates the coherence-aware design adopted in CoCo and highlights the importance of explicitly considering coherence-induced traffic during mapping and routing optimization. 

\subsection{Analysis of Cache Coherence Impact on Mapping}
\label{result_mapping}

In this section, we evaluate the impact of cache coherence on task mapping design, focusing on the benefits of incorporating coherence awareness into the mapping process. Specifically, the proposed Coherence-Guided Task Mapping (CGTM) is compared with (i) Tabu Search Mapping~\cite{b25}, which employs a metaheuristic search to iteratively refine task placements using memory-based move restrictions; (ii) Communication-aware Mapping~\cite{b11}, which places tasks based on communication intensity and congestion estimation; and (iii) HyDra~\cite{b308}, which performs runtime remapping to adapt to workload variations while preserving dependencies, to assess its effectiveness in improving communication efficiency under coherence-aware scenarios. These methods are originally designed for non-coherent environments. For a fair comparison, Tabu Search Mapping, Communication-aware Mapping, HyDra, and CGTM all employ deterministic XY routing under the same coherence-aware settings.


\begin{figure}[htbp]
    \centering
        \includegraphics[width=1.03\linewidth]{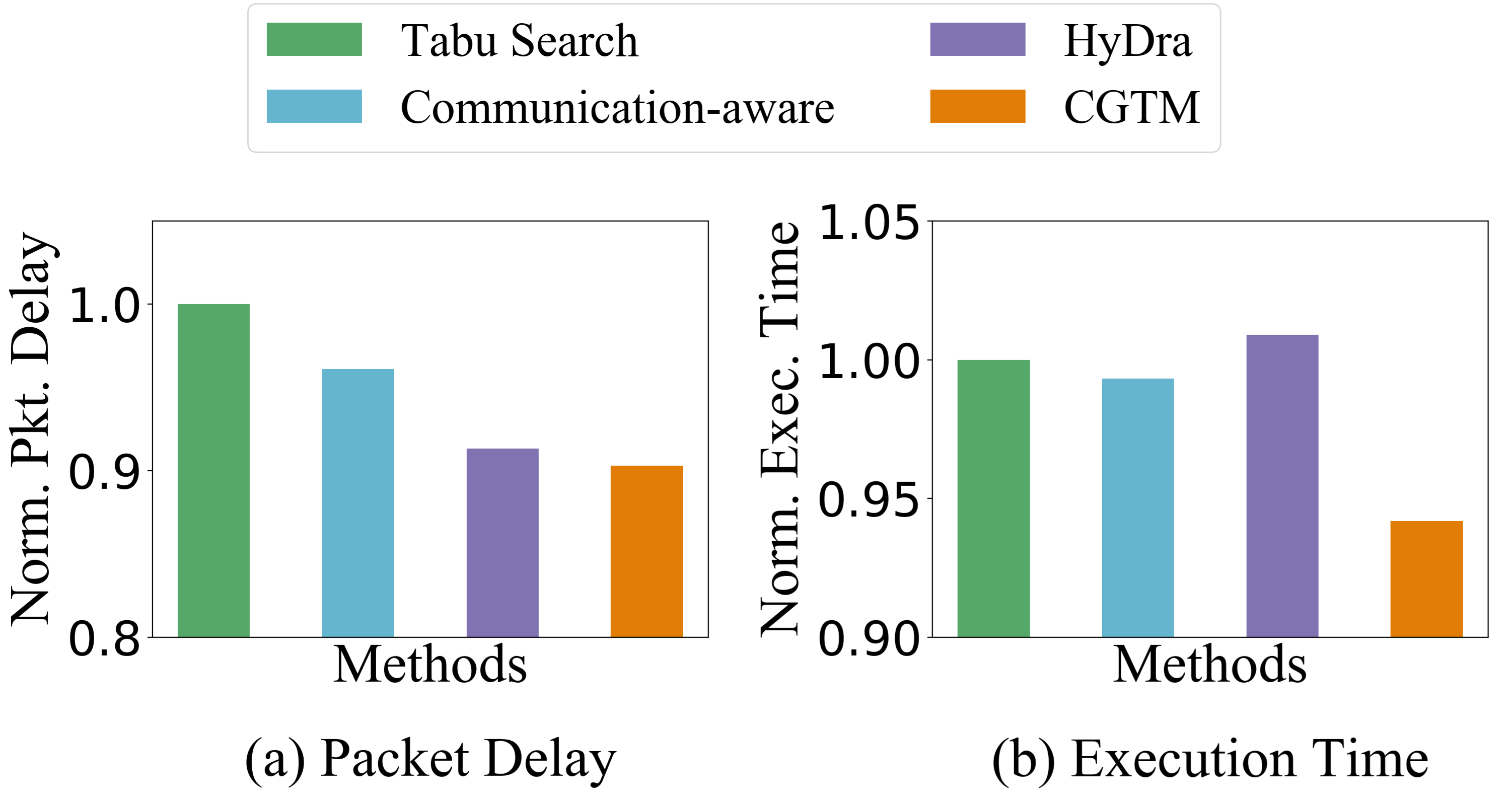}
   \caption{Mapping performance comparison across four methods}
    \label{fig:mapping_only}
\end{figure}

As shown in Fig.~\ref{fig:mapping_only}, CGTM achieves average packet-delay reductions of 9.69\%, 6.05\%, and 1.11\%, and execution-time reductions of 5.82\%, 5.18\%, and 6.66\%, compared with Tabu Search, Communication-aware Mapping, and HyDra, respectively. All methods were evaluated across ten independent runs, yielding coefficients of variation below 2.0\%, indicating stable and reliable performance improvements. These results further demonstrate the benefits of incorporating coherence awareness into task mapping for realistic systems.


Specifically, both Tabu Search and CGTM employ the same Tabu-based exploration through iterative task swaps, but differ in how candidate mappings are evaluated. Tabu Search relies on communication cost metrics that do not capture coherence-induced interactions. As a result, even optimized mappings may place coherence-intensive task pairs far apart or concentrate their communication on specific network regions, increasing hop count and traffic contention, which directly leads to higher packet delay. Similarly, communication-aware mapping optimizes placement based on communication intensity and congestion estimation, and HyDra adapts mappings to workload variations. However, neither explicitly models coherence-induced communication, and their optimization is driven by aggregate communication characteristics rather than the fine-grained and correlated nature of coherence interactions. Consequently, tasks with frequent coherence exchanges may still be placed far apart or mapped in a way that concentrates traffic on specific links, increasing both queuing delay and traversal latency. These delays in packet transmission further slow down coherence operations such as invalidations and cache-to-cache transfers, prolonging memory access and synchronization. As a result, packet delay accumulates along the critical path, ultimately increasing overall execution time. 

In contrast, CGTM explicitly incorporates coherence-aware information into the mapping process, enabling tasks with frequent coherence interactions to be placed in close proximity. This reduces communication distance and alleviates traffic concentration for coherence messages, leading to lower packet delay through fewer hops and reduced queuing time. The reduction in packet delay accelerates the delivery of coherence messages along the critical path, thereby decreasing memory access and synchronization stalls. As a result, overall execution time is reduced.

Overall, these results demonstrate that cache coherence plays a critical role in task mapping design, as ignoring coherence-induced interactions can lead to increased communication overhead and network contention. By explicitly incorporating coherence traffic characteristics, CGTM effectively improves communication efficiency and outperforms existing mapping approaches, highlighting its effectiveness in realistic coherence-aware systems.

\subsection{Performance Analysis of RLDWR}
\label{result_routing}

In this section, we evaluate the effectiveness of the proposed routing method, RLDWR. Four representative routing approaches are employed: (i) Communication-aware routing Dyad~\cite{b33}, which is congestion-adaptive and makes routing decisions based on queue lengths; (ii) RL-based routing~\cite{b29}, which learns adaptive routing policies from local network states (e.g., buffer occupancy and link utilization) to dynamically balance traffic and mitigate congestion; (iii) DRL-based routing~\cite{b30}, which models the network as a graph and leverages GNNs to capture structural relationships between nodes and links for improved routing efficiency; and (iv) DRL-based coherence-aware routing~\cite{b15}, which employs a neural network to learn link weights within an RL framework and incorporates a coherence-aware reward. To ensure a fair comparison, all methods are evaluated under the same fixed task mapping and coherence-aware scenarios. 

\begin{figure}[htbp]
    \centering
        \includegraphics[width=1.03\linewidth]{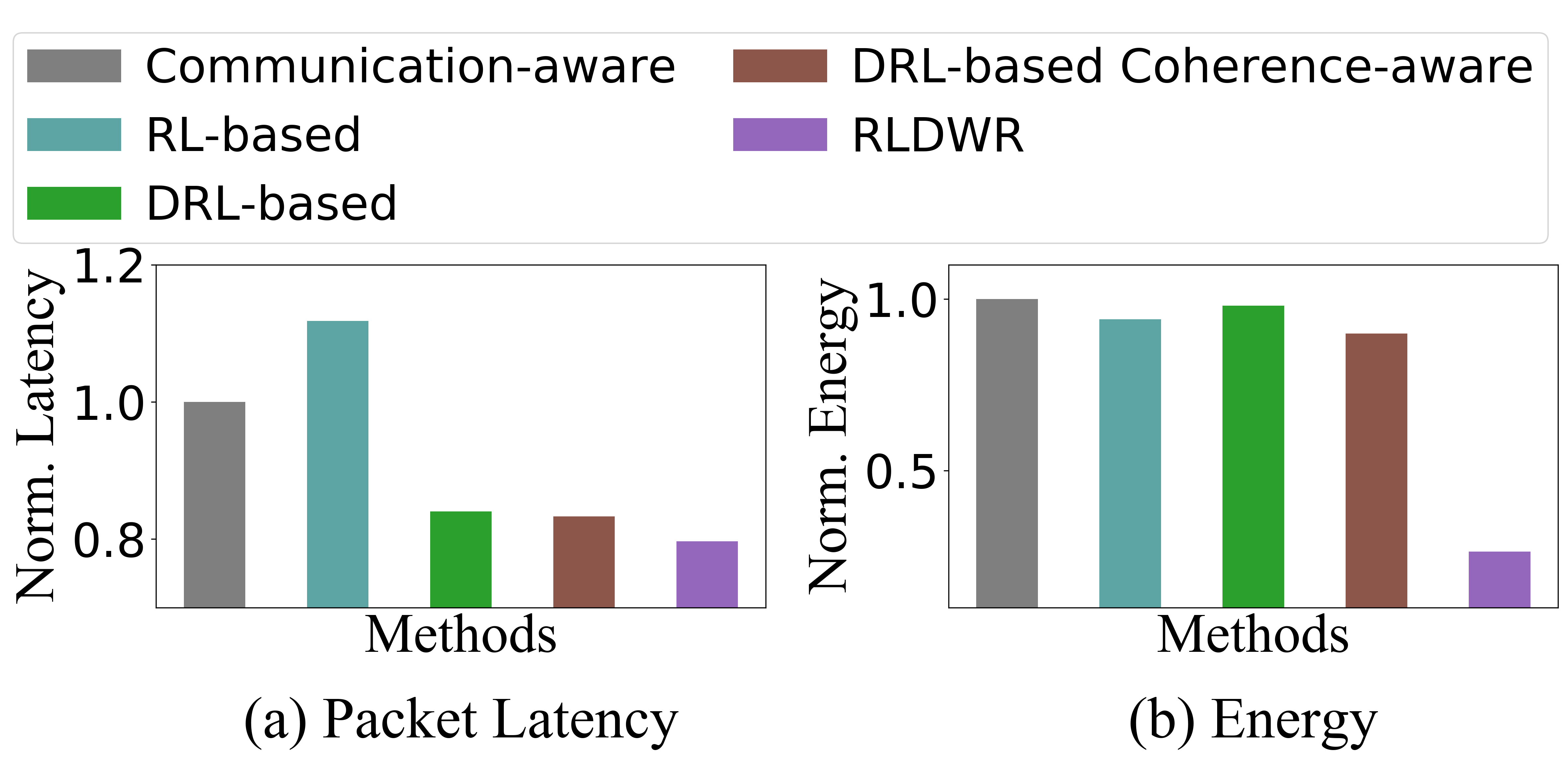}
   \caption{Routing performance comparison across five methods}
    \label{fig:routing_only}
\end{figure}

As shown in Fig.~\ref{fig:routing_only}, the proposed RLDWR achieves average packet latency reductions of 20.30\%, 28.73\%, 5.17\%, and 4.36\%, and total energy reductions of 73.68\%, 72.01\%, 73.15\%, and 70.71\%, compared to Communication-aware, RL-based, DRL-based, and DRL-based coherence-aware routing, respectively. These improvements stem from RLDWR’s ability to control path selection at a fine granularity while accounting for coherence-induced communication behavior, leading to reductions in both latency and energy consumption.

Although communication-aware, RL-based, and DRL-based routing approaches optimize path selection based on network conditions such as queue lengths, buffer occupancy, or learned traffic patterns, they do not explicitly capture coherence-induced communication characteristics. Consequently, while effective for general traffic balancing, these methods primarily react to emerging congestion and may exhibit suboptimal behavior under coherence-induced traffic patterns. In contrast, RLDWR incorporates coherence awareness into routing decisions by encoding directional preferences through optimized link weights based on observed communication behavior. This allows routing decisions to steer packets toward less contended directions, alleviating hotspot formation and reducing queue buildup. As a result, packets experience shorter waiting times at routers, leading to lower overall packet latency. Moreover, RLDWR reduces energy consumption by reducing unnecessary network activity. By discouraging traffic from persistently congested directions, packets spend less time buffered in routers, reducing buffer-access energy. In addition, by avoiding congested paths and limiting detours, RLDWR reduces the number of link traversals and switching operations required to deliver packets. This more efficient traffic distribution decreases both buffer and link activity, resulting in lower overall energy consumption.

Compared with DRL-based coherence-aware routing, which relies on topology-level decisions, RLDWR performs lightweight directional-level optimization. The coarser granularity of topology-level decisions limits control over traffic distribution across individual directions. Furthermore, although both approaches consider multiple optimization objectives, RLDWR adopts a unified cost model shared by the mapping and routing stages. As a result, routing decisions remain consistent with the criterion used during mapping evaluation while avoiding the additional complexity associated with maintaining separate objective representations. These design choices enable RLDWR to regulate traffic distribution more precisely under coherence-induced communication patterns. The resulting directional-weight configuration allows packets to avoid locally congested directions more effectively, reducing queue buildup and head-of-line blocking at routers, which in turn lowers queuing delay and overall latency. Furthermore, by minimizing unnecessary detours and reducing the time packets spend buffered in routers, RLDWR decreases both link switching activity and buffer accesses. This reduction in network activity lowers dynamic energy consumption, resulting in improved latency and reduced total energy consumption compared with DRL-based coherence-aware routing.

Overall, these results demonstrate that incorporating coherence awareness into routing is essential for effectively managing coherence-induced traffic in many-core systems. By directly optimizing a unified coherence-aware objective and employing optimized directional link weights, RLDWR achieves lower packet latency and reduced energy consumption compared to existing routing approaches. This highlights the effectiveness of RLDWR in improving routing efficiency under realistic coherence-aware scenarios.

\subsection{Performance Analysis of CoCo}
\label{result_coco}
In this section, we first conduct a sensitivity analysis of the unified cost model parameters and analyze the benefits of jointly optimizing task mapping and routing by comparing CGTM-only, RLDWR-only, and the complete CoCo framework. We then compare CoCo with state-of-the-art approaches, including SMART, SMT, and MARCO, to evaluate its effectiveness in terms of NoC efficiency, coherence overhead, and overall system performance. To gain further insight into the observed performance improvements, we analyze the resulting communication patterns. Finally, we evaluate the scalability of CoCo and its sensitivity to different input datasets. 




\textbf{Sensitivity Analysis of Unified Cost Model Parameters.}
As discussed in Section~\ref{cost}, the unified cost model contains four coefficients, namely $\alpha$, $\beta$, $\lambda_1$, and $\lambda_2$, which control the relative importance of communication path cost, coherence overhead, load imbalance, and directional imbalance, respectively. To assess the sensitivity of these parameters, each coefficient is varied within $\{0.2, 0.5, 0.8, 1.0, 1.2\}$ while the remaining coefficients are held constant. Since all cost components are scaled before aggregation, the selected range provides a broad variation in the relative emphasis assigned to each optimization objective while avoiding extreme weight settings that could cause a single objective to dominate the unified cost. This enables a systematic evaluation of parameter sensitivity under balanced optimization conditions.

\begin{figure}[htbp]
    \centering
    
    \begin{subfigure}[b]{0.47\textwidth}
        \centering
        \includegraphics[width=\linewidth]{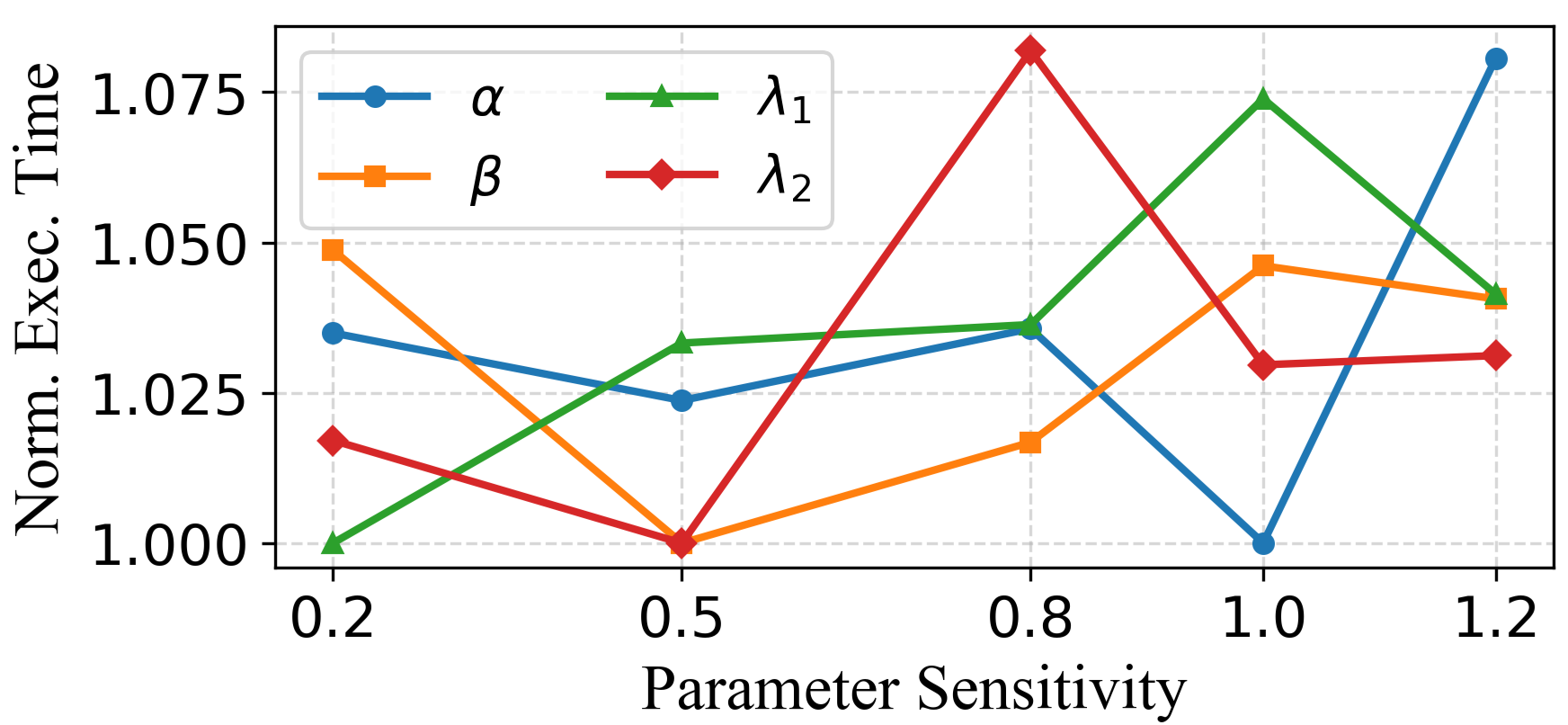}
        \caption{Sensitivity to unified cost model coefficients}
        \label{fig:sensitivity}
    \end{subfigure}

    \begin{subfigure}[b]{0.48\textwidth}
        \centering
        \includegraphics[width=\linewidth]{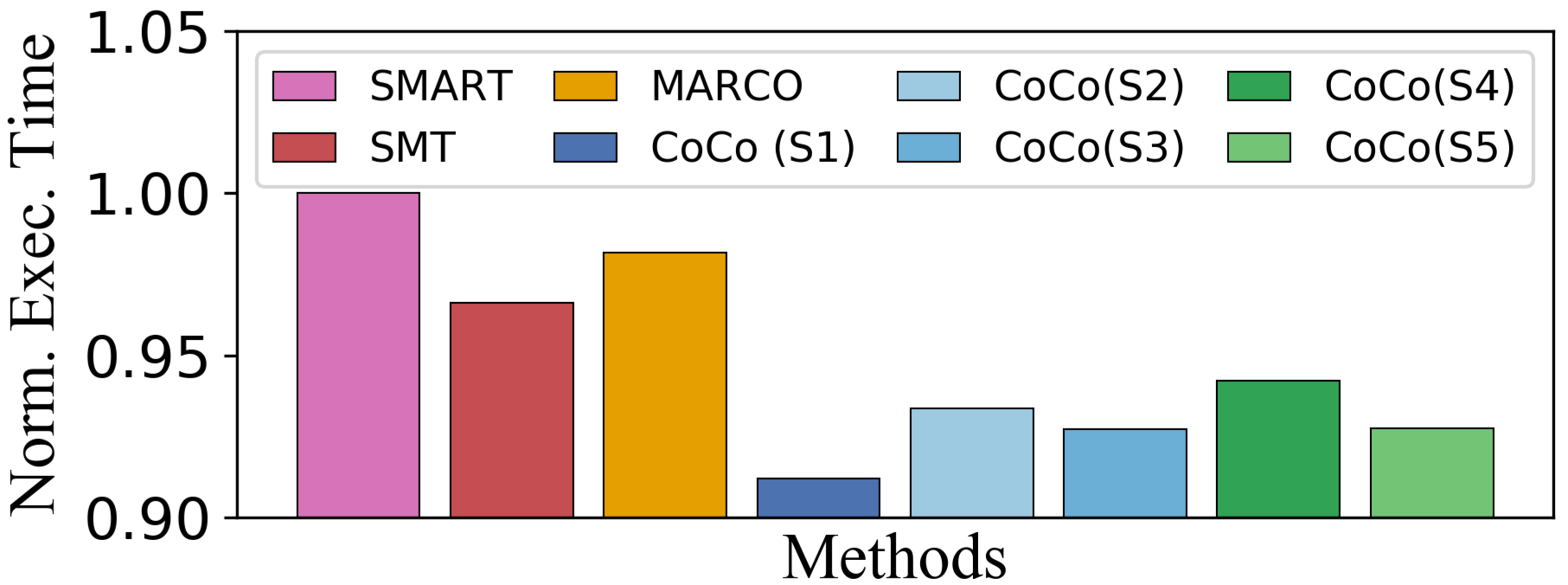}
        \caption{Robustness across different coefficient configurations}
        \label{fig:robustness}
    \end{subfigure}
      
    \caption{Normalized execution time under varying cost-model coefficients: (a) sensitivity to coefficient values, and (b) robustness across different coefficient configurations.}
    \label{fig:sen}
\end{figure}

As shown in Fig.~\ref{fig:sensitivity}, the best-performing values are $\alpha=1.0$, $\beta=0.5$, $\lambda_1=0.2$, and $\lambda_2=0.5$. These values are combined to form the configuration S1. Compared with the other evaluated values of each coefficient, the selected parameter values reduce execution time by up to 7.46\%, 4.65\%, 6.88\%, and 7.57\% for $\alpha$, $\beta$, $\lambda_1$, and $\lambda_2$, respectively. These results indicate that the coefficients have a noticeable impact on optimization effectiveness. Values that are either too small or too large generally lead to higher execution time, suggesting that effective optimization requires a balanced weighting of the objectives within the unified cost model.

To assess the robustness of CoCo under moderate parameter variations, Fig.~\ref{fig:robustness} compares the optimal configuration S1 with four perturbed settings: S2 ($\alpha=0.5$, $\beta=0.5$, $\lambda_1=0.2$, and $\lambda_2=0.5$), S3 ($\alpha=1.0$, $\beta=0.8$, $\lambda_1=0.2$, and $\lambda_2=0.5$), S4 ($\alpha=1.0$, $\beta=0.5$, $\lambda_1=0.5$, and $\lambda_2=0.5$), and S5 ($\alpha=1.0$, $\beta=0.5$, $\lambda_1=0.2$, and $\lambda_2=0.2$). Although S1 achieves the shortest execution time, all evaluated configurations consistently outperform SMART, SMT, and MARCO, reducing execution time by 2.50\% to 8.80\%. This indicates that CoCo remains effective across a range of parameter settings rather than relying on a narrowly tuned configuration. The results therefore demonstrate the robustness of the proposed framework to moderate parameter variations, highlighting the ability of the unified coherence-aware cost model to consistently coordinate mapping and routing decisions across different parameter settings. Accordingly, S1 is adopted as the default configuration in the remaining experiments.




\textbf{Enhancement of Co-optimization.} Based on the insights from Section~\ref{result_mapping} and Section~\ref{result_routing}, both CGTM and RLDWR achieve significant performance gains compared to existing methods. We further analyze the impact of integrating these two components to highlight the importance of co-optimization. In this evaluation, CoCo is compared with CGTM-only and RLDWR-only, with all three methods evaluated under the same coherence-aware environment.

\begin{figure}[htbp]
    \centering
        \includegraphics[width=1.03\linewidth]{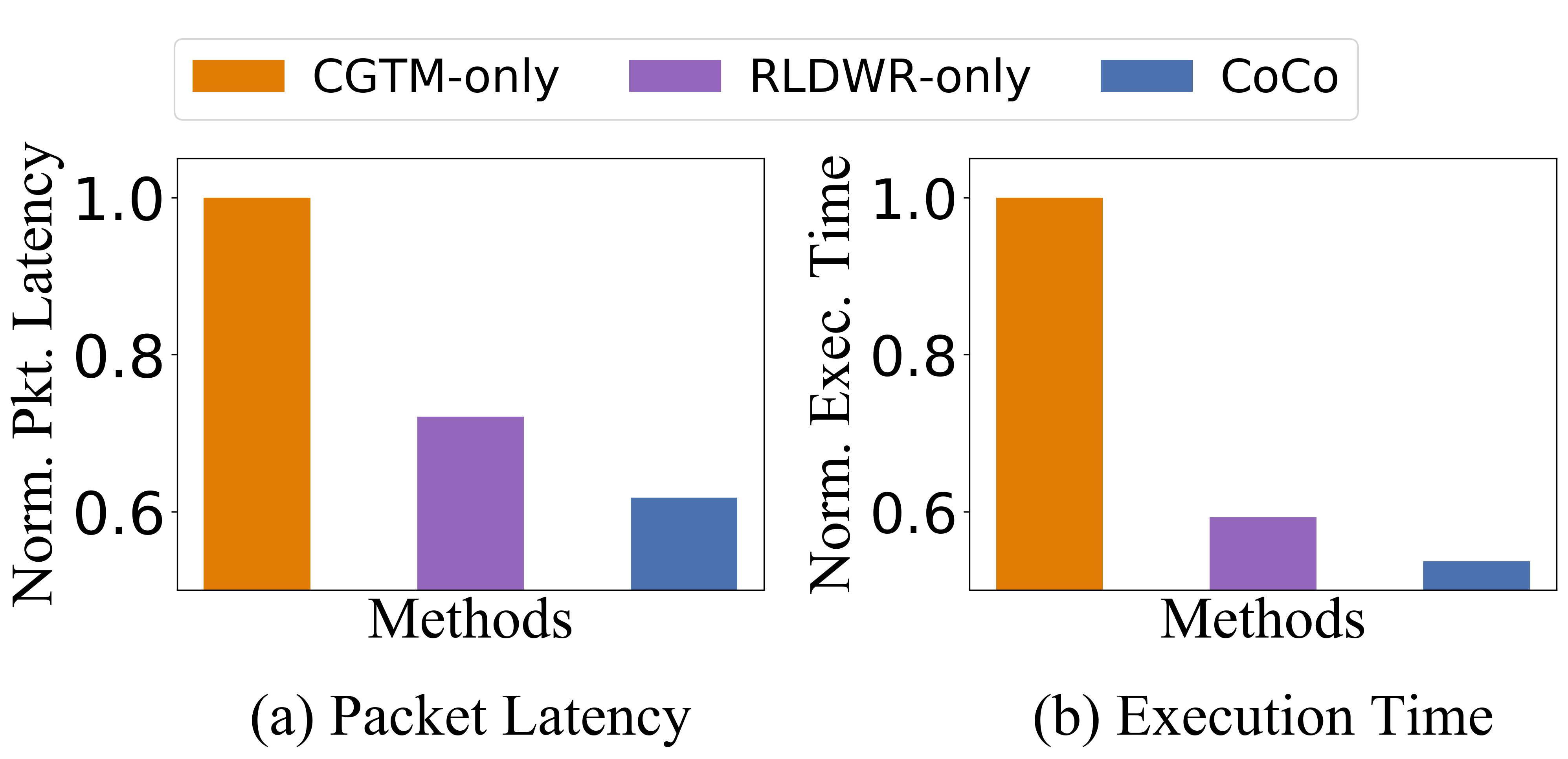}
   \caption{Performance comparison of CGTM-only, RLDWR-only and CoCo.}
    \label{fig:coco}
\end{figure}

As shown in Fig.~\ref{fig:coco}, CoCo achieves average packet latency reductions of 38.2\% and 14.3\%, and total execution time reductions of 46.3\% and 9.4\%, compared to CGTM-only and RLDWR-only, respectively. These results indicate that, although each component individually provides substantial improvements, their integration is essential to fully exploit the benefits of coherence-aware optimization, leading to superior overall performance. This advantage stems from the complementary interaction between coherence-aware task mapping and routing. While CGTM reduces communication overhead by placing frequently interacting tasks closer together, it cannot eliminate congestion under deterministic routing, leading to higher packet latency. Similarly, RLDWR improves traffic distribution through optimized directional routing preferences, but its effectiveness is constrained by the initial task placement, which may still incur long-distance communication and contention. By integrating CGTM and RLDWR under a unified coherence-aware cost model, CoCo jointly optimizes communication locality and traffic distribution. CGTM reduces the volume and distance of coherence-induced communication, while RLDWR employs optimized directional link weights to guide traffic toward less contended directions. This coordination reduces both communication distance and contention, allowing packets to traverse fewer hops and experience shorter queues, thereby lowering packet latency. At the system level, these improvements reduce memory-access and synchronization stalls, leading to significant execution-time reductions compared with CGTM-only and RLDWR-only.


Overall, these results demonstrate that co-optimization of task mapping and routing is essential for effectively handling coherence-induced communication in many-core systems. By integrating CGTM and RLDWR under a unified coherence-aware cost model, CoCo reduces communication distance and balances traffic, leading to lower packet latency and fewer stalls, thereby improving total execution time.

\begin{figure}[htbp]
    \centering
    
    \begin{subfigure}[b]{0.50\textwidth}
        \centering
        \includegraphics[width=\linewidth]{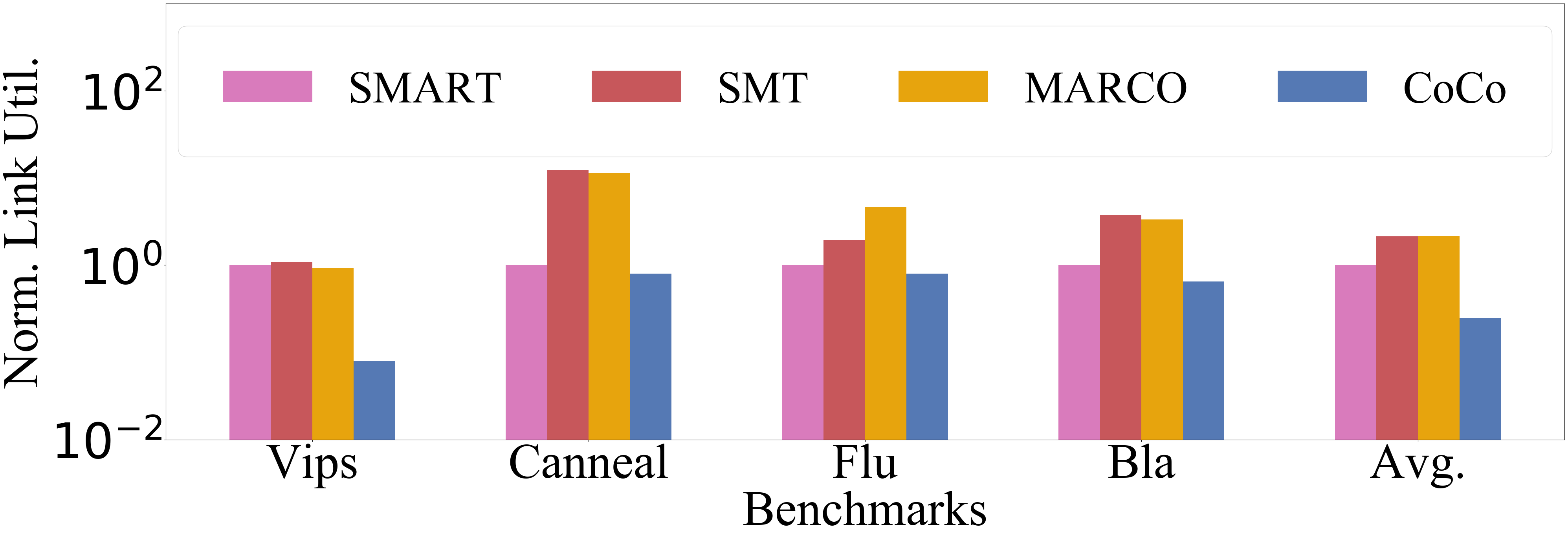}
        \caption{Average link utilization}
        \label{fig:16_link}
    \end{subfigure}

    \begin{subfigure}[b]{0.50\textwidth}
        \centering
        \includegraphics[width=\linewidth]{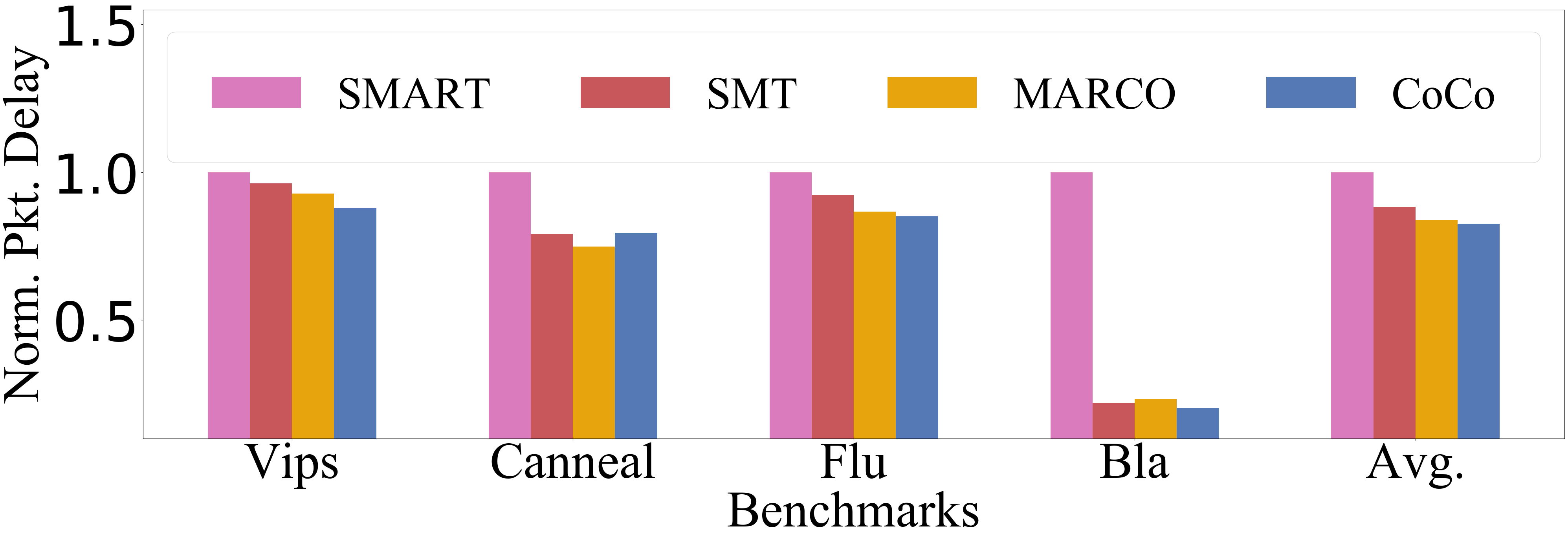}
        \caption{Average Packet Delay}
        \label{fig:16_delay}
    \end{subfigure}
      
    \caption{Normalized performance across methods: (a)Average link utilization, and (b) Average packet delay.}
    \label{fig:noc}
\end{figure}

\textbf{Enhancement of NoC Efficiency.}
As illustrated in Fig.~\ref{fig:noc}, our method CoCo achieves average link utilization reductions of 75.14\%, 88.41\%, and 88.46\%, and average packet delay reductions of 17.40\%, 6.40\%, and 1.56\%, compared to SMART, SMT and MARCO, respectively.

From the perspective of link utilization, SMART employs power-gating–aware task mapping but relies on fixed routing paths during execution, while SMT constructs contention-free mappings with diversified routing paths to reduce potential conflicts. In parallel, MARCO adopts a communication-aware mapping and routing strategy over flexible interconnects, capturing general communication patterns to improve traffic distribution. However, these approaches do not explicitly account for coherence-induced communication. As a result, under realistic workloads, bursty and correlated coherence messages may repeatedly concentrate traffic on specific links, leading to high link utilization and persistent congestion. In contrast, CoCo incorporates coherence awareness into both mapping and routing through a unified cost model. At the mapping stage, CGTM reduces the spatial concentration of coherence-induced communication by placing frequently interacting tasks closer together, thereby lowering the traffic intensity on heavily used links. Building on this, RLDWR optimizes directional link weights to guide routing decisions according to observed communication behavior, enabling packets to be distributed more efficiently under coherence-induced traffic patterns. This joint optimization reduces the likelihood that multiple packets compete for the same set of links, promoting a more balanced traffic distribution across the network.
As a result, link usage becomes more evenly balanced across the network rather than being concentrated on a few hotspots, thereby reducing persistent congestion and limiting queue buildup at routers. Consequently, traffic is distributed more efficiently across available communication paths, reducing excessive activity on heavily utilized links. This leads to more efficient utilization of network resources and ultimately lower overall link utilization.


From the perspective of packet delay, SMART and SMT rely on fixed routing decisions or design-time optimization and do not explicitly account for coherence-induced communication patterns. As a result, routing decisions fail to effectively accommodate the irregular and highly correlated traffic generated by coherence operations, where multiple messages frequently target the same directions within short time intervals. Packets therefore tend to repeatedly traverse heavily utilized links, leading to localized congestion and preventing traffic from being effectively redistributed across the network. This results in sustained queue buildup at intermediate routers, increasing both queuing and traversal delays and ultimately leading to higher overall packet delay. In contrast, CoCo reduces packet delay through coordinated optimization of communication locality and traffic distribution. By leveraging coherence-aware mapping, CoCo places frequently interacting tasks closer together, thereby reducing long-distance communication and mitigating traffic concentration across distant regions of the network. In addition, RLDWR utilizes optimized directional link weights to reduce traffic concentration on heavily utilized directions, promoting a more balanced communication pattern across the network. Together, these mechanisms reduce sustained queue buildup at heavily loaded routers while shortening communication distances. Consequently, packets experience both shorter travel paths and less waiting time at intermediate routers, resulting in lower hop-by-hop latency, shorter queuing times, and ultimately reduced overall packet delay. 
Compared with MARCO, CoCo still achieves lower overall packet delay because it explicitly incorporates coherence-induced communication into both the mapping and routing stages. This enables CoCo to better capture coherence-related communication behavior and reduce contention under coherence-intensive workloads. However, the overall delay reduction is relatively modest, primarily due to the behavior observed in the Canneal benchmark, where MARCO achieves a 5.84\% lower packet delay than CoCo. Specifically, Canneal exhibits irregular communication patterns with weaker locality and more dispersed coherence traffic, resulting in fewer persistent congestion hotspots. Consequently, the benefits of coherence-aware optimization are less pronounced, as the communication behavior is less structured and provides fewer opportunities to reduce traffic concentration through coherence-aware mapping and directional-weight optimization. In contrast, MARCO's communication-aware mapping and flexible routing strategy are better suited to irregular traffic patterns, resulting in lower packet delay on Canneal. Consequently, the overall packet-delay advantage of CoCo over MARCO is reduced. 

In addition, for the Blackscholes (Bla) benchmark, the packet delay achieved by SMT, MARCO, and CoCo is significantly lower than that of SMART, with reductions of 77.99\%, 76.64\%, and 79.75\%, respectively. This is because Bla exhibits relatively regular communication patterns with stronger locality, where optimized task placement already reduces communication distance effectively. Under such conditions, deterministic routing in SMART becomes a limiting factor, as it cannot alleviate residual contention along fixed paths. In contrast, SMT, MARCO, and CoCo employ routing mechanisms that provide greater flexibility in traffic distribution, enabling communication to be spread more effectively across the network and reducing congestion. As a result, they achieve substantially lower packet delay than SMART.

Overall, CoCo effectively reduces link utilization and packet delay by jointly optimizing communication locality and traffic distribution under coherence-aware scenarios. While its performance gains are less pronounced for irregular workloads such as Canneal, it consistently outperforms existing approaches in structured and coherence-intensive scenarios, where methods like SMART are limited by static routing. These results highlight the importance of coherence-aware co-optimization and the impact of workload characteristics on NoC efficiency.

\begin{figure}[htbp]
    \centering
        \includegraphics[width=\linewidth]{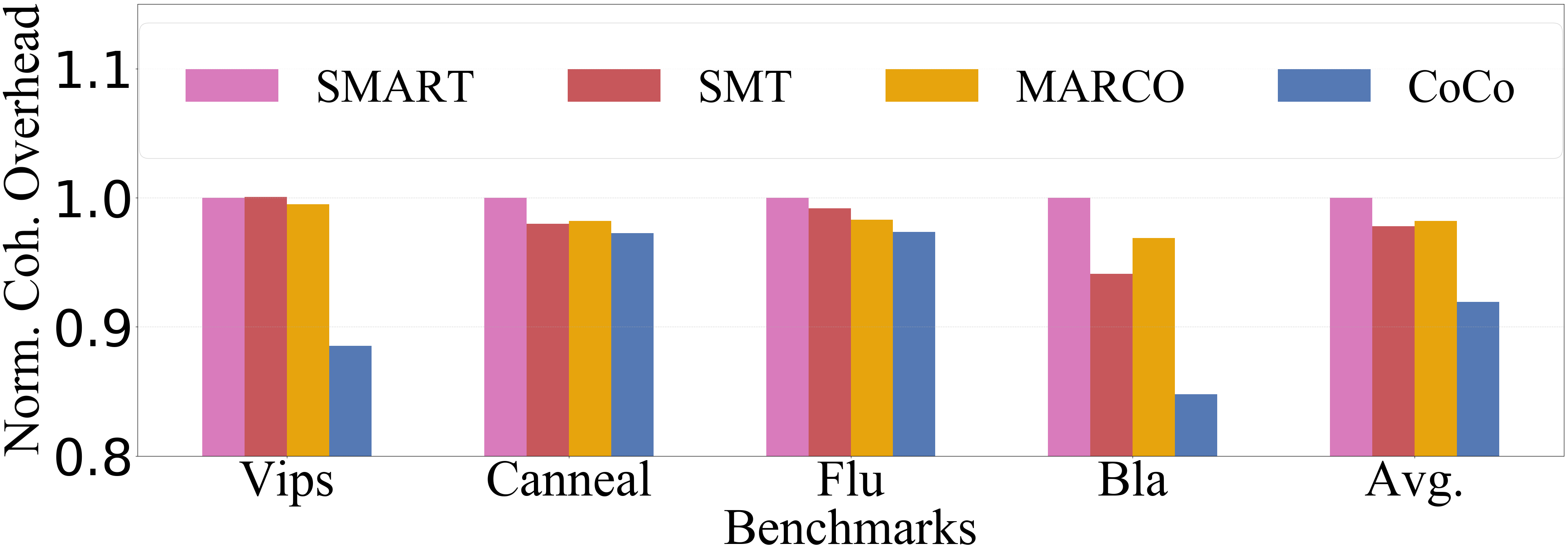}
   \caption{Normalized coherence overhead across four methods}
    \label{fig:16_coh_time}
\end{figure}

\textbf{Enhancement of Coherence Overhead.}
As illustrated in Fig.~\ref{fig:16_coh_time}, CoCo achieves total coherence overhead reductions of 8.08\%, 6.00\%, and 6.04\% compared to SMART, SMT, and MARCO, respectively.
SMART improves task placement but relies on deterministic routing, forcing coherence messages to traverse minimal paths. Under bursty coherence traffic, where invalidations and cache-to-cache transfers repeatedly target the same destinations, these fixed paths become persistent hotspots, increasing queuing delays and coherence latency. While SMT attempts to reduce contention by generating diversified paths during mapping, these paths are predetermined and cannot adapt to variations in coherence traffic patterns, leading to imbalance and localized congestion over time. MARCO adopts communication-aware mapping and routing strategies based on aggregate communication characteristics, but it does not distinguish coherence-specific traffic patterns. As a result, it cannot effectively capture the repeated and correlated nature of coherence messages, allowing traffic to accumulate on certain links and increasing coherence overhead.

In contrast, CoCo explicitly incorporates coherence-induced communication into both mapping and routing under a unified cost model. At the mapping stage, CGTM evaluates task placement by incorporating coherence interaction intensity and path cost (Section~\ref {cost}), placing tasks with frequent coherence exchanges closer together. This reduces the hop count of coherence messages and limits their traversal across multiple routers, thereby lowering coherence overhead. At the routing stage, RLDWR optimizes directional link weights according to coherence-aware communication patterns, including the frequency and directionality of coherence traffic. Guided by the unified cost model, the optimization process learns routing preferences that distribute coherence-related communication more evenly across routing directions. As a result, packets are more likely to utilize routing directions favored by the optimized weight configuration, reducing traffic concentration on heavily utilized links. This mechanism alleviates queue buildup and head-of-line blocking at routers, thereby mitigating network contention. Consequently, coherence messages experience shorter traversal and waiting times, reducing packet stalls and decreasing the time spent handling coherence traffic.


Overall, CoCo effectively reduces coherence overhead across diverse benchmarks, demonstrating robust performance under varying coherence communication patterns.

\begin{figure}[htbp]
    \centering
    
    \begin{subfigure}[b]{0.50\textwidth}
        \centering
        \includegraphics[width=\linewidth]{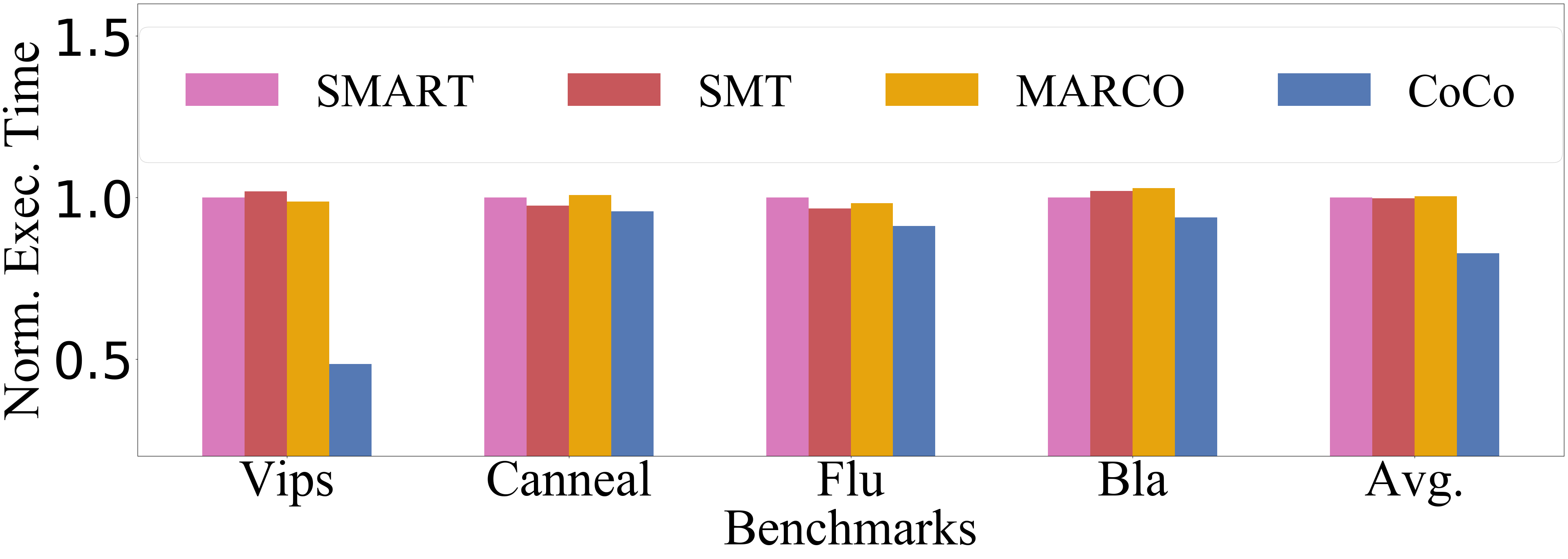}
        \caption{Execution time}
        \label{fig:16_exe_time}
    \end{subfigure}

    \begin{subfigure}[b]{0.50\textwidth}
        \centering
        \includegraphics[width=\linewidth]{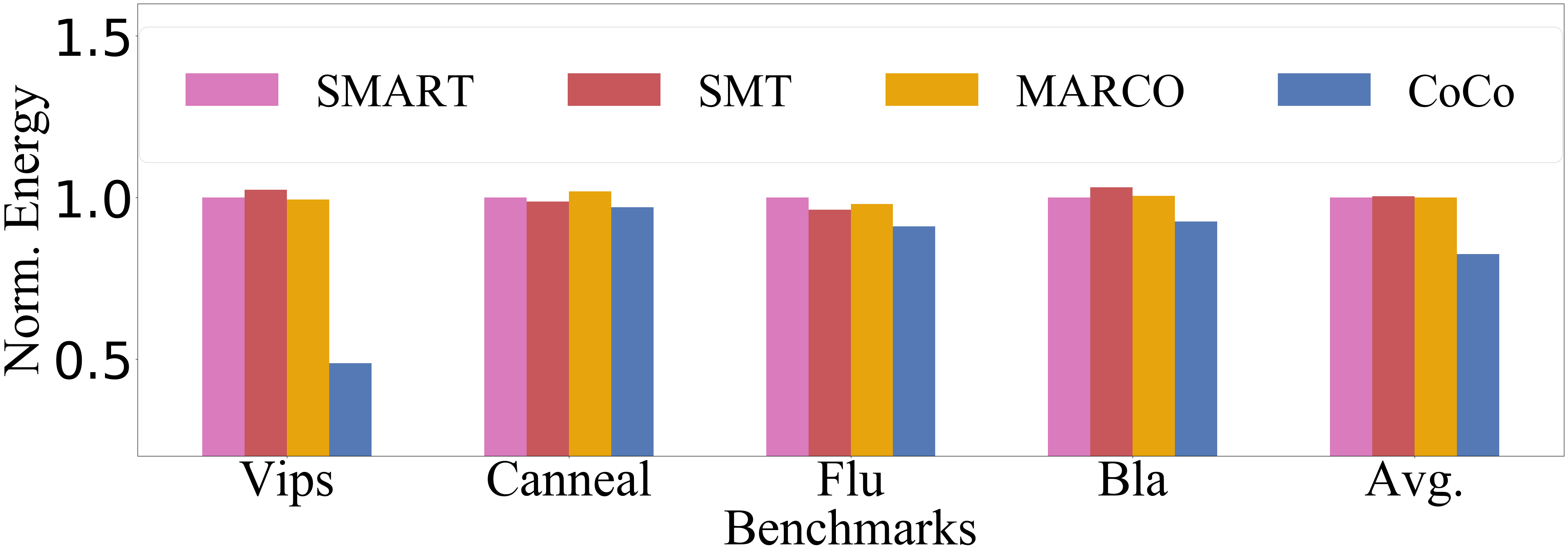}
        \caption{Total energy}
        \label{fig:16_energy}
    \end{subfigure}
      
    \caption{Normalized performance across methods: (a) Total execution time, and (b) Total energy.}
    \label{fig:system}
\end{figure} 

\textbf{Enhancement of System Performance.}
As shown in Fig.~\ref{fig:system}, CoCo achieves total execution time reductions of 17.30\%, 17.08\%, and 17.58\%, and total energy savings of 17.51\%, 17.85\%, and 17.51\% compared to SMART, SMT, and MARCO, respectively. These gains arise from improved NoC efficiency and reduced coherence overhead.

SMART adopts a fixed routing scheme in which packets follow predetermined minimal paths, without adapting to runtime traffic conditions. Under coherence-intensive workloads, where coherence messages are repeatedly exchanged among a subset of nodes, this leads to persistent congestion, increasing queuing delay, traversal latency, and execution time. The resulting congestion also raises buffer occupancy and router activity, leading to higher buffer and link energy use. While SMT improves task placement and introduces path diversity at design time, its routing remains static during execution. As coherence traffic patterns vary, predefined paths may become suboptimal, causing uneven traffic distribution. This imbalance increases packet waiting time and may lead to longer routing paths, contributing to higher communication delay and execution time. Increased hop counts and buffering further elevate link activity and buffer accesses, resulting in higher energy consumption. Meanwhile, MARCO integrates communication-aware mapping with flexible routing over the interconnect, but the use of separate cost evaluators weakens coordination between mapping and routing decisions. This can produce task placements that generate concentrated coherence traffic, limiting the ability of routing to distribute traffic efficiently and thereby increasing congestion. Consequently, both queuing delay and traversal latency increase. Moreover, inefficient coordination may cause unnecessary data movement and redundant link usage, further increasing energy consumption.

In contrast, CoCo achieves coordinated optimization of communication locality and traffic distribution through a unified cost model. By aligning task placement with routing optimization, it reduces the likelihood of long-distance or highly concentrated communication patterns while preventing traffic from accumulating on a small set of links. This coordination ensures that communication demand is both reduced and more evenly distributed across the network. As a result, packets traverse fewer hops and experience shorter waiting times at intermediate routers, reducing both traversal latency and queuing delay. These improvements shorten communication stalls on the critical path, leading to reduced execution time.

The observed energy savings can be attributed to reductions in both network activity and execution time. By reducing communication distance, traffic concentration, and coherence-induced communication overhead, CoCo decreases link switching activity, router utilization, and buffer usage, thereby lowering the energy consumed by packet transmission and buffering. In addition, the shorter execution time further reduces overall energy consumption, resulting in the energy savings observed in Fig.~\ref{fig:16_energy}.


Overall, CoCo effectively reduces communication overhead and enhances both execution time and energy efficiency under realistic workload scenarios by coordinating coherence-aware task placement and routing decisions within a unified model.

\begin{figure}[htbp]
    \centering
        \includegraphics[width=0.9\linewidth]{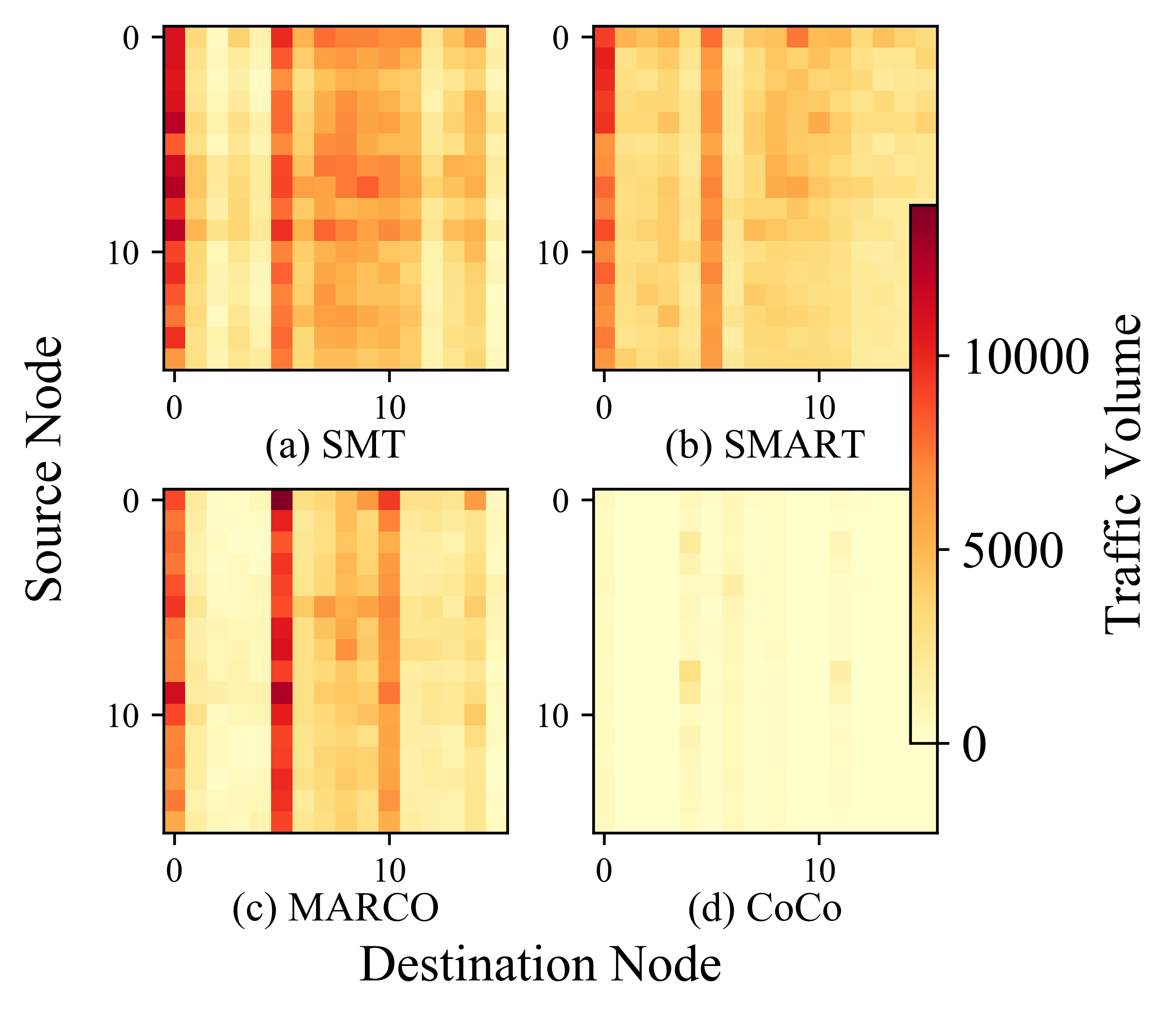}
   \caption{Communication traffic heatmaps for SMT, SMART, MARCO, and CoCo. Darker colors indicate higher communication volumes between source and destination nodes.}
    \label{fig:heatmap}
\end{figure}

\textbf{Communication Pattern Analysis.}
To further investigate the sources of the performance improvements achieved by CoCo, we analyze communication patterns using traffic heatmaps. Fig.~\ref{fig:heatmap} illustrates the traffic volume between source and destination nodes for SMT, SMART, MARCO, and CoCo on a $4\times4$-core system, where darker colors indicate higher traffic volumes. As shown in Fig.~\ref{fig:heatmap}(a)-(c), SMT, SMART, and MARCO exhibit several pronounced high-traffic regions, indicating that communication is concentrated among a limited set of source-destination pairs. Such traffic concentration can increase network contention and coherence-related communication overhead. In contrast, Fig.~\ref{fig:heatmap}(d) shows that CoCo substantially reduces the intensity and frequency of these high-traffic regions, resulting in a noticeably more balanced communication pattern. The resulting traffic distribution provides insight into the performance improvements reported in Section~\ref{result_coco}. By reducing hotspot formation and the associated network contention, CoCo lowers packet delay, link utilization, and coherence overhead, which collectively contribute to the observed reductions in execution time and energy consumption.



\begin{figure}[htbp]
    \centering
        \includegraphics[width=1.05\linewidth]{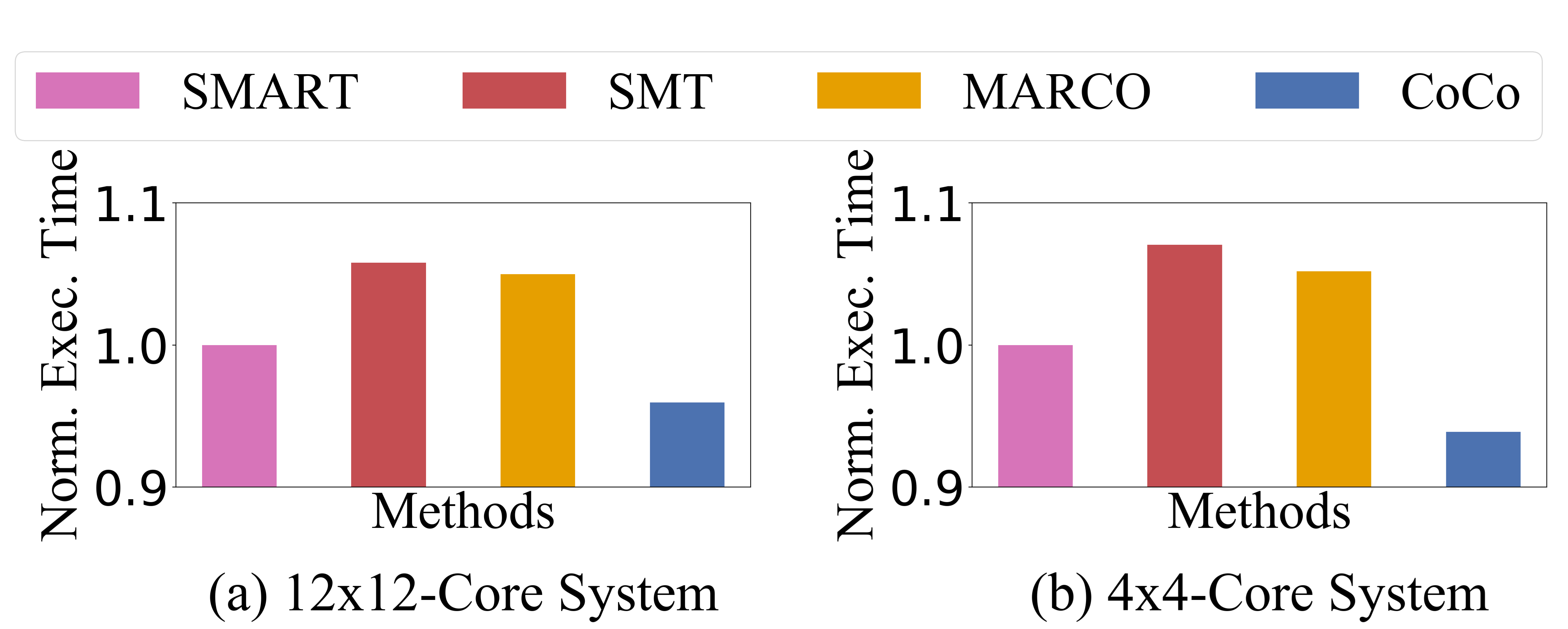}
   \caption{Normalized execution time across four methods on $4\times4$ and $12\times 12$-core system.}
    \label{fig:scale}
\end{figure}

\textbf{Scalability Analysis.} To evaluate scalability, we assess CoCo on $4\times4$ and $12\times12$ mesh systems. As shown in TABLE~\ref{tab:scale}, the design-time cost remains stable across different system sizes, with variations within 3\% relative to the $8\times8$ system. Fig.~\ref{fig:scale} presents the corresponding performance results. Compared with SMART, SMT, and MARCO, CoCo reduces execution time by 4.04\%, 9.29\%, and 8.59\%, respectively, on the $4\times4$ system, and by 6.12\%, 12.29\%, and 10.74\%, respectively, on the $12\times12$ system. These results demonstrate that CoCo maintains stable design-time cost while consistently improving performance across different system scales, highlighting the scalability and broad applicability of the proposed coherence-aware mapping–routing co-optimization framework to many-core systems with varying architectural configurations.


\begin{table}[htbp]
\centering
\caption{Design-time reduction across different scales}
\label{tab:scale}
\small
\renewcommand{\arraystretch}{1.15}
\begin{tabular}{|p{3.00cm}|p{4.20cm}|}
\hline
\textbf{System Scale} & \textbf{Design-Time Reduction} \\ \hline
4 $\times$ 4   & 1.84\% \\ \hline
8 $\times$ 8   & 2.60\% \\ \hline
12 $\times$ 12 & 0 (reference) \\ \hline
\end{tabular}
\end{table}

\textbf{Sensitivity Analysis of Input Datasets.}
To evaluate the sensitivity of CoCo to input dataset variations, we conduct experiments using three Canneal datasets: small, medium, and large. The same optimization framework is applied across all datasets. As shown in TABLE~\ref{tab:diff_size}, the execution-time variation remains below 1\% across the evaluated datasets, with deviations of only 0.894\% and 0.144\% for the medium and large datasets, respectively, relative to the small dataset. These results indicate that CoCo maintains stable optimization effectiveness despite variations in workload characteristics. By jointly optimizing task mapping and routing through a unified coherence-aware cost model, CoCo avoids over-specialization to dataset-specific communication patterns. Consequently, the framework exhibits strong generalization across different input datasets while maintaining consistently stable optimization performance.



\begin{table}[htbp]
\centering
\caption{Execution-time variation across different datasets}
\label{tab:diff_size}
\small
\renewcommand{\arraystretch}{1.15}
\begin{tabular}{|p{3.00cm}|p{4.20cm}|}
\hline
\textbf{Input Dataset} & \textbf{Execution-Time Reduction} \\ \hline
Canneal-Small   & 0 (reference) \\ \hline
Canneal-Medium  & 0.894\% \\ \hline
Canneal-Large   & 0.144\% \\ \hline
\end{tabular}
\end{table}

\section{Conclusion}
This work shows that coherence behavior strongly influences the joint optimization of task mapping and routing, and thus significantly impacts overall application execution. However, most existing approaches typically overlook cache coherence, leaving a substantial portion of coherence-induced communication unaccounted for and thereby leading to a mismatch between optimization objectives and actual communication patterns. Moreover, these approaches typically employ separate cost evaluators for mapping and routing, which complicates the coordination of multiple optimization objectives and fails to capture the coherence-induced interdependence between the two. To address these challenges, we propose CoCo, a coherence-aware co-optimization framework that jointly integrates task mapping and routing under a unified cost model. The model incorporates communication cost, coherence overhead, and load imbalance into a single objective, enabling effective trade-offs and coherence-aware optimization. Guided by this model, CoCo combines coherence-guided task mapping with reinforcement learning-based routing, which adjusts directional link weights according to observed communication behavior. This tightly coupled design captures the interdependence between coherence, mapping, and routing, enabling more effective optimization for realistic many-core applications. Experimental results demonstrate that CoCo reduces link utilization by 88.46\%, packet delay by 17.40\%, and execution time by 17.58\% compared to existing approaches, underscoring the importance of coherence-aware co-optimization for many-core systems.

\bibliographystyle{IEEEtran}

\bibliography{refs}

\newpage

\vfill

\end{document}